\documentclass[showpacs,aps,graphicx,twocolumn,pra,reprint]{revtex4}
\usepackage{graphicx}
\usepackage{verbatim}
\usepackage[caption=false,font=scriptsize,labelfont=sf,textfont=sf]{subfig}

\usepackage{amsmath}
\graphicspath{{./pic/}}
\usepackage{multirow}
\usepackage{amssymb}
\usepackage[T1]{fontenc}

\begin{document}

\title{Efficient multipartite entanglement purification with non-identical states}

\author{Hao Qin$^{1,2}$, Ming-Ming Du$^{1}$, Xi-Yun Li$^{2}$, Wei Zhong$^{3}$, Lan Zhou$^{2}$\footnote{Email address:zhoul@njupt.edu.cn}, Yu-Bo Sheng$^{1,3}$\footnote{Email address:shengyb@njupt.edu.cn}}
\address{
$^1$College of Electronic and Optical Engineering $\&$ College of Flexible Electronics (Future Technology), Nanjing University of Posts and Telecommunications, Nanjing 210023, China\\
$^2$School of Science, Nanjing University of Posts and Telecommunications, Nanjing,
210023, China\\
$^3$Institute of Quantum Information and Technology, Nanjing University of Posts and Telecommunications, Nanjing 210003, China\\
}
\date{\today }

\begin{abstract}
We present an efficient and general multipartite entanglement purification protocol (MEPP) for N-photon systems in Greenberger-Horne-Zeilinger (GHZ) states with non-identical input states. As a branch of entanglement purification, besides the cases of successful purification, the recurrence MEPP actually has the reusable discarded items which are usually regarded as a failure.
Our protocol contains two parts for bit-flip error correction. The first one is the conventional MEPP, corresponding successful cases. The second one includes two efficient approaches, recycling purification with entanglement link and direct residual entanglement purification, that can utilize discarded items. 
We also make a comparison between two approaches. Which method to use depends on initial input states, and in most cases the approach of direct residual purification is optimal for it not only may obtain a higher fidelity entangled state but also it does not require additional sophisticated links. In addition, for phase-flip errors, the discarded items still have available residual entanglement in the case of different input states. 
With these approaches, this MEPP has a higher efficiency than all previous MEPPs and it may have potential applications in the future long-distance quantum communications and networks.

\end{abstract}
\maketitle

\section{Introduction}
As a special quantum correlation property, quantum entanglement is widely used in quantum information processing protcols, such as quantum teleportation \cite{qt}, quantum key distribution (QKD)  \cite{qkd1,qkd2}, quantum secure direct communication (QSDC) \cite{QSDC1,QSDC2,QSDC3,QSDC4,QSDC5,QSDC6}, distribued secure quantum machine learning \cite{DSQML}
and quantum repeaters \cite{repeater1}. In addition, quantum secret sharing (QSS) \cite{qss1,qss2,qss3} and quantum state sharing (QST) \cite{qsts1,qsts2,qsts3} require maximally  multipartite entangled states. On the other hand, the powerful acceleration of quantum computation depends on multipartite entanglement \cite{quantumcomputation}.  However, the maximally entangled states are inevitably  degraded  by noise during the entanglement distribution and storage.   The degraded entanglement will decrease the security of quantum communication protocols. 

Entanglement purification is the approach to distill the high quality entanglement from low quality entanglement \cite{repp1,diffepp1,diffepp2,repp2,repp3,repp4,earlymepp1,earlymepp2,earlymepp3,rmepp4,rmepp5,depp1,depp2,depp3,dmepp1,36-pra-deng-mepp,dmepp2,mbepp1,mbepp2,mbepp3,optimization-epp2,dur,13-oe-diffepp,mbmepp1,mbmepp2,hybrid, mbmepp3,experiment1,experiment2,experiment3,experiment4,experiment5,experiment6,experiment7,experiment8,review},  and it becomes a key element in quantum repeaters \cite{repeater1}. The concept of entanglement purification was proposed by Bennett \textit{et al}. in 1996 \cite{repp1}.  Deustch \textit{et al.} improved the first entanglement purification protocol (EPP) using unitary operations \cite{diffepp1}. In 2001, Pan\textit{ et al.} proposed the first EPP using feasible linear optics \cite{repp2}. In 2008, the efficient EPP  based on parametric down-conversion sources with cross-Kerr nonlinearity was proposed \cite{repp4}. In 2013, the concept of measurement-based entanglement purification was proposed and the measurement-based entanglement purification was extended to the logic-qubit entanglement \cite{mbepp1,mbepp2,mbepp3,mbmepp1,mbmepp2,mbmepp3}. Some fundamental entanglement purification experiments were realized in optics \cite{experiment1,experiment2,experiment3,experiment4,experiment5}, atom\cite{experiment6}, solid state\cite{experiment7}, superconduct systems\cite{experiment8}. 

For multipartite systems, Murao 
et al. proposed the first multipartite entanglement purification protocol (MEPP) in 1998 \cite{earlymepp1}. This protocol was also based on controlled-not (CNOT) gate following similar ideas in Ref. \cite{repp1}. In 2008, using the cross-Kerr nonlinearity to construct the quantum nondemolition detector (QND), Sheng et al. proposed the MEPP for polarization entanglement in optical system \cite{earlymepp2}.  In 2011, Deng proposed an efficient MEPP for N-photon systems in Greenbeerger-Horne-Zeilinger (GHZ) states with QNDs \cite{36-pra-deng-mepp}. The innovation of this MEPP is that the discarded instances in the conventional MEPPs can be used to regenerate high-fidelity N-photon entangled systems with entanglement link.
 
So far, existing  MEPPs usually consider the initial mixed states with the same fidelity. 
Actually, like the two-particle entanglement, multipartite entangled systems are from different ensembles in a practical application. For instance, in quantum repeaters with multiplexed memory \cite{quantum-repeaters-with-multiplexed-memory1,quantum-repeaters-with-multiplexed-memory2}, the entangled states are stored in different quantum memories. Therefore, it is natural that the fidelity of the entangled states are different with different memory time. In this paper, we present an efficient recurrence MEPP for N-photon systems in GHZ states with different initial fidelities. It contains the following parts. In the case of bit-flip errors,
we show that the discarded items in conventional MEPPs can be reused by two different approaches. The first is the approach of entanglement link, called scheme P1. 
By measuring some photons, the discarded items can become entangled N$^\prime$-photon (2 $\le$ N$^\prime < $ N) subsystems which can be reused to generate a N-photon entangled state similar to Ref. \cite{36-pra-deng-mepp}. 
The second approach named P1$^\prime$ is the direct residual entanglement purification. We show that when initial N-photon systems are in the mixed GHZ states with different fidelities, such discarded items still have entanglement which can be reused directly to improve the yield of high quality entangled states like Ref. \cite{13-oe-diffepp}. 
We also conduct specific analyses and comparison of the approach of residual entanglement purification and entanglement link.
Likewise, after performing the purification for the phase-flip error correction, the discarded items still have entanglement which can also be reused to improve the yield in a second purification round. Our MEPP is firstly proposed for a three-photon system and then extended to arbitrary N-photon systems. 

This paper is organized as follows. In Sec. \ref{P1}, we first take three-photon systems in GHZ states as an example to discuss scheme P1 for bit-flip errors in the general case of different ensembles. 
In the scheme of P1, we can use the approach of entanglement link to improve the fidelity for the discarded items. 
Second, we propose another new scheme P1$^\prime$ utilizing the discarded instances to deal with bit-flip errors in Sec. \ref{P1'}. We show that the discarded items still have entanglement and they can be reused directly. Then, in Sec. \ref{N-bit}, we give the description of N-photon EPP for bit-flip errors. The three-photon MEPP for phase-flip error in the case of different ensembles is described and extended to the N-photon case in Sec. \ref{phase}. After that, in Sec. \ref{GOMEPP} we make a discussion and compare schemes P1 and  P1$^\prime$ in detail, and our general optimized MEPP for the case of different initial ensembles is proposed. Finally, we present a conclusion in Sec. \ref{dc}.

\section{Multipartite entanglement purification for bit-flip errors with non-identical states}\label{bit}
Before starting this MEPP, we first introduce the key element of this protocol, 
that is, the QND constructed by cross-Kerr nonlinearity.
QND here has the functions of a photon-number detector (PND) and a parity-check gate by measuring the phase shift of the coherent state. The parity-check gate can be used to distinguish superpositions and mixtures of the even parity (e) states $|HH\rangle ,|VV\rangle $ from the odd parity (o) states $|HV\rangle ,|VH\rangle $, and allows us to circumvent the Knill bounds \cite{PhysRevA.68.064303} and construct a near deterministic CNOT gate \cite{qnd-kerr}.
Such QNDs were wildly used in entanglement purification \cite{repp4,depp1,dmepp1}. Here $|H\rangle$ and $|V\rangle$ present horizontal and vertical polarization of the photon.

In order to describe the details of MEPP, we first consider the case of three-photon entanglement in a GHZ state. For three-photon GHZ state, there are eight GHZ states considering the polarization degree of freedom, which can be written as follows:
\begin{align}\label{eq1}
{|\Phi_0^\pm\rangle }_{ABC}=\frac{1}{\sqrt{2} } {(|HHH\rangle \pm|VVV\rangle )}_{ABC},\nonumber\\
{|\Phi_1^\pm\rangle }_{ABC}=\frac{1}{\sqrt{2} } {(|VHH\rangle \pm|HVV\rangle )}_{ABC},\nonumber\\
{|\Phi_2^\pm\rangle }_{ABC}=\frac{1}{\sqrt{2} } {(|HVH\rangle \pm|VHV\rangle )}_{ABC},\nonumber\\
{|\Phi_3^\pm\rangle }_{ABC}=\frac{1}{\sqrt{2} } {(|HHV\rangle \pm|VVH\rangle )}_{ABC}.
\end{align}
The subscripts $A$, $B$ and $C$ here represent photons sent to Alice, Bob and Charlie, respectively. Assuming that the original GHZ state transmitted among the three parties is ${|\Phi_0^+\rangle }_{ABC}$. The three-photon system in the states ${|\Phi_1^+\rangle }_{ABC}$, ${|\Phi_2^+\rangle}_{ABC}$ and ${|\Phi_3^+\rangle }_{ABC}$ respectively indicates that  a bit-flip error takes place on the first qubit, the second qubit and the third qubit, respectively. If ${|\Phi_0^+\rangle }_{ABC}$ becomes ${|\Phi_0^-\rangle }_{ABC}$, there is a phase-flip error. Sometimes, a bit-flip error and a phase-flip error will take place simultaneously. The purpose of purifying the three-photon entangled systems  requires correcting both bit-flip errors and phase-flip errors. We first discuss the principle of MEPP for purifying the bit-flip errors here, marking the part as P1/P1$^\prime$ of MEPP, and discuss the P2 part of MEPP for phase-flip errors in the next section.

In general situation, the mixed states are different and come from different ensembles. Suppose Alice, Bob and Charlie share the source ensemble ${^1\!}\rho$ and target ensemble ${^2\!}\rho$ as
\begin{align}\label{eq2}
{^1\!}{\rho}_{ABC}=&{^1\!}F_0|\Phi_0^+\rangle \langle \Phi_0^+|+{^1\!}F_1|\Phi_1^+\rangle \langle \Phi_1^+|\nonumber\\
&+{^1\!}F_2|\Phi_2^+\rangle \langle \Phi_2^+|+{^1\!}F_3|\Phi_3^+\rangle \langle \Phi_3^+|,\nonumber\\
{^2\!}{\rho}_{ABC}=&{^2\!}F_0|\Phi_0^+\rangle \langle \Phi_0^+|+{^2\!}F_1|\Phi_1^+\rangle \langle \Phi_1^+|\nonumber\\
&+{^2\!}F_2|\Phi_2^+\rangle \langle \Phi_2^+|+{^2\!}F_3|\Phi_3^+\rangle \langle \Phi_3^+|.
\end{align}
Here ${^\alpha\!}F_0\!=\!\langle \Phi_0^+|{^\alpha\!}\rho|\Phi_0^+\rangle $ $(\alpha\!\in\!\left\{1,2\right\})$, ${^\alpha\!}F_0$ is the fidelity of the mixed states, and notice the normalization condition $ {\textstyle \sum_{i}} $  
${^\alpha\!}F_i\!=\!1(i\!\in\!\left\{0,1,2,3\right\})$. We suppose ${^1\!}F_0\!>\! {^2\!}F_0$ for simplicity. The density matrix ${^\alpha\!}\rho$  means that there is a bit-flip error on the first qubit, the second qubit, and the third qubit with a probability of ${^\alpha\!}F_1$, ${^\alpha\!}F_2$ and ${^\alpha\!}F_3$, respectively. Then, we first generalize the protocol in Ref. \cite{36-pra-deng-mepp} and describe the P1 part in the case of different ensembles.

\subsection{MEPP-P1: correction for bit-flip errors in the entanglement purification protocol}\label{P1}
In order to obtain some high-fidelity three-photon entangled states, the three parties first divide their subsystems in the source ensemble ${^1}\rho$ and target ensemble ${^2}\rho$ into many groups, and each group composed of a pair of three-photon entangled states. We label each group with the three-photon state is $A_1B_1C_1$ and  $A_2B_2C_2$, respectively. 

\begin{figure}[ht]
\includegraphics[width=.5\textwidth]{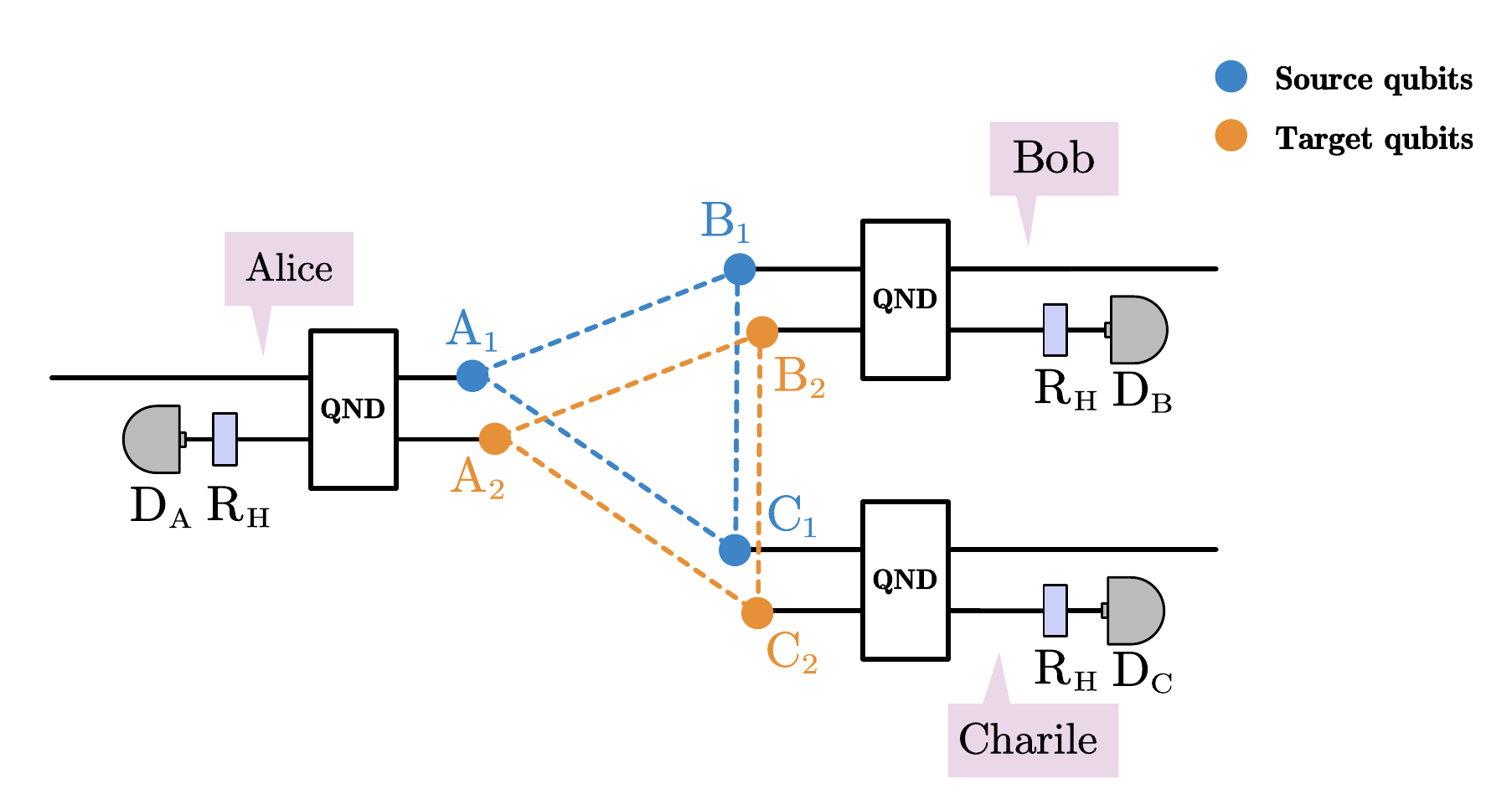}
\caption{The principle of our conventional three-photon entanglement purification protocol for bit-flip errors with QNDs \cite{36-pra-deng-mepp}. The wave plate $R_H$ represents a Hadamard operation and it is used to transform the polarization states $|H\rangle $ and $|V\rangle $ into $|+\rangle\! =\!\frac{1}{\sqrt{2}}(|H\rangle \!+\!|V\rangle )$ and $|-\rangle \!=\!\frac{1}{\sqrt{2}}(|H\rangle \!-\!|V\rangle )$, respectively. $D_A$, $D_B$ and $D_C$ represent the single-photon measurements with the basis $Z\!=\!\left \{|H\rangle ,|V\rangle \right \}$ done by Alice, Bob, and Charlie, respectively. The blue and orange dots represent the qubits of the source systems and the target systems respectively, and the dotted lines represent the long-distance quantum channels.}
\label{pic1}
\end{figure}
The principle of MEPP to correct the bit-flip errors for three-photon GHZ state is shown in Fig. \ref{pic1}.
Two pairs of initial state  $A_1B_1C_1$ and $A_2B_2C_2$ can be regarded as the mixture of the 16 pure states, namely, ${|\Phi_i^+\rangle }_1\otimes{|\Phi_j^+\rangle }_2$ with the probability of ${^1\!}F_i{^2\!}F_j$ $(i,j\!\in\!\left\{0,1,2,3\right\})$. As shown in Fig. \ref{pic1}, for each group, Alice, Bob and Charlie let the two photons pass through the QNDs in the up and down spatial mode to perform the parity check, respectively. 
After the QNDs, there are $2^3$ parity-check results. If they all obtain an even parity or an odd parity (i.e. eee/ooo), these instances correspond to the identity combinations ${|\Phi_i^+\rangle }_1\otimes{|\Phi_i^+\rangle }_2$ $(i\!\in\!\left\{0,1,2,3\right\})$, just similar to previous MEPPs \cite{earlymepp1,earlymepp2,earlymepp3}. Otherwise, those instances will correspond to the cross combinations ${|\Phi_i^+\rangle }_1\otimes{|\Phi_j^+\rangle }_2$ $(i\!\ne \!j\!\in\!\left\{0,1,2,3\right\})$.

\subsubsection{Conventional entanglement purification with identity combinations}
If  the three parties all obtain an even parity (eee) after the QNDs, the quantum system $A_1B_1C_1A_2B_2C_2$ is in a new mixed state which is composed of the four states
\begin{subequations}
\begin{align}
{|\phi_0\rangle }_{ABC}=&\frac{1}{\sqrt{2}}({|HHH\rangle }_{A_1B_1C_1}{|HHH\rangle }_{A_2B_2C_2}\nonumber\\
&+{|VVV\rangle }_{A_1B_1C_1}{|VVV\rangle }_{A_2B_2C_2}),
\label{eq3-a} \\
{|\phi_1\rangle }_{ABC}=&\frac{1}{\sqrt{2}}({|VHH\rangle }_{A_1B_1C_1}{|VHH\rangle }_{A_2B_2C_2}\nonumber\\
&+{|HVV\rangle }_{A_1B_1C_1}{|HVV\rangle }_{A_2B_2C_2}),
\label{eq3b} \\
{|\phi_2\rangle }_{ABC}=&\frac{1}{\sqrt{2}}({|HVH\rangle }_{A_1B_1C_1}{|HVH\rangle }_{A_2B_2C_2}\nonumber\\
&+{|VHV\rangle }_{A_1B_1C_1}{|VHV\rangle }_{A_2B_2C_2}),
\label{eq3c} \\
{|\phi_3\rangle }_{ABC}=&\frac{1}{\sqrt{2}}({|HHV\rangle }_{A_1B_1C_1}{|HHV\rangle }_{A_2B_2C_2}\nonumber\\
&+{|VVH\rangle }_{A_1B_1C_1}{|VVH\rangle }_{A_2B_2C_2}),
\label{eq3d} 
\end{align}
\end{subequations}
with a probability of $\frac{1}{2}{^1\!}F_0{^2\!}F_0$, $\frac{1}{2}{^1\!}F_1{^2\!}F_1$, $\frac{1}{2}{^1\!}F_2{^2\!}F_2$ and $\frac{1}{2}{^1\!}F_3{^2\!}F_3$, respectively.
In addition, when the three parties all obtain an odd parity (ooo) after the QNDs, the quantum system is in another mixed state which is composed of the four states
\begin{subequations}
\begin{align}
{|\psi_0\rangle }_{ABC}=&\frac{1}{\sqrt{2}}({|HHH\rangle }_{A_1B_1C_1}{|VVV\rangle }_{A_2B_2C_2}\nonumber\\
&+{|VVV\rangle }_{A_1B_1C_1}{|HHH\rangle }_{A_2B_2C_2}),
\label{eq4-a} \\
{|\psi_1\rangle }_{ABC}=&\frac{1}{\sqrt{2}}({|VHH\rangle }_{A_1B_1C_1}{|HVV\rangle }_{A_2B_2C_2}\nonumber\\
&+{|HVV\rangle }_{A_1B_1C_1}{|VHH\rangle }_{A_2B_2C_2}),
\label{eq4b} \\
{|\psi_2\rangle }_{ABC}=&\frac{1}{\sqrt{2}}({|HVH\rangle }_{A_1B_1C_1}{|VHV\rangle }_{A_2B_2C_2}\nonumber\\
&+{|VHV\rangle }_{A_1B_1C_1}{|HVH\rangle }_{A_2B_2C_2}),
\label{eq4c} \\
{|\psi_3\rangle }_{ABC}=&\frac{1}{\sqrt{2}}({|HHV\rangle }_{A_1B_1C_1}{|VVH\rangle }_{A_2B_2C_2}\nonumber\\
&+{|VVH\rangle }_{A_1B_1C_1}{|HHV\rangle }_{A_2B_2C_2}),
\label{eq4d} 
\end{align}
\end{subequations}
with a probability of $\frac{1}{2}{^1\!}F_0{^2\!}F_0$, $\frac{1}{2}{^1\!}F_1{^2\!}F_1$, $\frac{1}{2}{^1\!}F_2{^2\!}F_2$ and $\frac{1}{2}{^1\!}F_3{^2\!}F_3$, respectively.

For the two parity-check results (eee/ooo), after the Hadamard operation on each photon in the down spatial mode, Alice, Bob, and Charlie measure the photons $A_2B_2C_2$ with the basis $Z\!=\!\left \{|H\rangle ,|V\rangle \right \}$ in Fig. \ref{pic1}. The outcomes will divide the cases into two groups according to the parity of $|V\rangle $ 
 number. In the first group, the number of the outcomes $|V\rangle $ is even, then they obtain the states $|\Phi_0^+\rangle $, $|\Phi_1^+\rangle $, $|\Phi_2^+\rangle $ and $|\Phi_3^+\rangle $ with the probabilities $\frac{1}{4}{^1\!}F_0{^2\!}F_0$, $\frac{1}{4}{^1\!}F_1{^2\!}F_1$, $\frac{1}{4}{^1\!}F_2{^2\!}F_2$ and $\frac{1}{4}{^1\!}F_3{^2\!}F_3$. In another group,  the number of $|V\rangle $ is odd, the parties obtain the states $|\Phi_0^-\rangle $, $|\Phi_1^-\rangle $, $|\Phi_2^-\rangle $ and $|\Phi_3^-\rangle $ with the probabilities $\frac{1}{4}{^1\!}F_0{^2\!}F_0$, $\frac{1}{4}{^1\!}F_1{^2\!}F_1$, $\frac{1}{4}{^1\!}F_2{^2\!}F_2$ and $\frac{1}{4}{^1\!}F_3{^2\!}F_3$. Obviously, they can transform the state $|\Phi_i^-\rangle $ into the state $|\Phi_i^+\rangle $ with a phase-flip operation ${\sigma}_z\!=\!|H\rangle \langle H|\!-\!|V\rangle \langle V|$ on any qubit of the source state. 

Therefore, after passing through the QNDs, for the same parity-check results (eee/ooo) corresponding to the identity combinations ${|\Phi_i^+\rangle }_1\otimes{|\Phi_i^+\rangle }_2\ $, they measure the photons through the down spatial modes and obtain a new ensemble in the state
\begin{align}\label{eq5}
{\rho}_{ABC}'=&{F}_0'|\Phi_0^+\rangle \langle \Phi_0^+|+{F}_1'|\Phi_1^+\rangle \langle \Phi_1^+|\nonumber\\
&+{F}_2'|\Phi_2^+\rangle \langle \Phi_2^+|+{F}_3'|\Phi_3^+\rangle \langle \Phi_3^+|,
\end{align}
with
\begin{align}\label{eq6}
F'_0=\frac{{^1\!}F_0{^2\!}F_0}{{^1\!}F_0{^2\!}F_0+{^1\!}F_1{^2\!}F_1+{^1\!}F_2{^2\!}F_2+{^1\!}F_3{^2\!}F_3},\nonumber\\
F_1'=\frac{{^1\!}F_1{^2\!}F_1}{{^1\!}F_0{^2\!}F_0+{^1\!}F_1{^2\!}F_1+{^1\!}F_2{^2\!}F_2+{^1\!}F_3{^2\!}F_3},\nonumber\\
F_2'=\frac{{^1\!}F_2{^2\!}F_2}{{^1\!}F_0{^2\!}F_0+{^1\!}F_1{^2\!}F_1+{^1\!}F_2{^2\!}F_2+{^1\!}F_3{^2\!}F_3},\nonumber\\
F_3'=\frac{{^1\!}F_3{^2\!}F_3}{{^1\!}F_0{^2\!}F_0+{^1\!}F_1{^2\!}F_1+{^1\!}F_2{^2\!}F_2+{^1\!}F_3{^2\!}F_3}.
\end{align}
The fidelity of the new system  $F'_0> {^1\!}F_0$ if ${^2\!}F_0$ satisfies the relation
\begin{align}\label{eq7}
{^2\!}F_0\!>\!\frac{{^1\!}F_1{^2\!}F_1\!+\!{^1\!}F_2{^2\!}F_2\!+\!(1-{^1\!}F_0\!-\!{^1\!}F_1\!-\!{^1\!}F_2)(1\!-\!{^2\!}F_1\!-\!{^2\!}F_2)}{2-2{^1\!}F_0-{^1\!}F_1-{^1\!}F_2}.
\end{align}
Eq. $\eqref{eq7}$ is the criterion for successfully purifying and obtaining a new system with higher fidelity from the identity combinations. Apparently, if ${^1\!}F_0\!=\!{^2\!}F_0$, then revert to the special situation under the same ensemble hypothesis \cite{36-pra-deng-mepp}.

Consider the special case of the symmetric noise channels, then the initial three-photon entangled system before being purified is symmetric, namely, ${^\alpha\!}F_i\!=\!\frac{1-{^\alpha\!}F_0}{3}$ $(\alpha\!\in\!\left\{1,2\right\}$, $i\!\in\!\left\{1,2,3\right\})$, 
and Eq. $\eqref{eq7}$ will be simplified to
\begin{align}\label{eq8}
{^2\!}F_0> \frac{1}{4}.
\end{align}
Thus, $F'_0\!>\! {^1\!}F_0$ if ${^1\!}F_0\!>\! {^2\!}F_0\!>\! 1/4$. It's similar to the case of two-particle purification from the different ensembles \cite{13-oe-diffepp}. That is, the lower bound of ${^2\!}F_0$ is $1/4$, which is completely consistent with the entanglement purification condition under the assumptions of the same ensemble and the symmetric entanglement system \cite{36-pra-deng-mepp}, and it will revert to the specific case when ${^1\!}F_0\!=\!{^2\!}F_0$. 

\subsubsection{Recycling entanglement purification with cross combinations using entanglement link }
\begin{table*}
\centering  
\caption{The states of two-photon system obtained from cross combinations and their probabilities.}  
\label{table1}
\resizebox{\linewidth}{!}{
\begin{tabular}{|c||c|c||c|c||c|c|}
\hline

Parities & \multicolumn{2}{c||}{eoe/oeo} &\multicolumn{2}{c||}{oee/eoo} &\multicolumn{2}{c|}{eeo/ooe}   \\ \hline

\multirow{2}{*}{Cross combinations} & $|\Phi_0^+\rangle \otimes|\Phi_2^+\rangle $ & $|\Phi_1^+\rangle \otimes|\Phi_3^+\rangle $ & $|\Phi_0^+\rangle \otimes|\Phi_1^+\rangle $ & $|\Phi_2^+\rangle \otimes|\Phi_3^+\rangle $ & $|\Phi_0^+\rangle \otimes|\Phi_3^+\rangle $ & $|\Phi_1^+\rangle \otimes|\Phi_2^+\rangle $ 
\\  \cline{2-7}
& $|\Phi_2^+\rangle \otimes|\Phi_0^+\rangle $ & $|\Phi_3^+\rangle \otimes|\Phi_1^+\rangle $ & $|\Phi_1^+\rangle \otimes|\Phi_0^+\rangle $ & $|\Phi_3^+\rangle \otimes|\Phi_2^+\rangle $ & $|\Phi_3^+\rangle \otimes|\Phi_0^+\rangle $ & $|\Phi_2^+\rangle \otimes|\Phi_1^+\rangle $ \\  \hline

Probabilities & ${^1\!}F_0{^2\!}F_2+{^1\!}F_2{^2\!}F_0$ & ${^1\!}F_1{^2\!}F_3+{^1\!}F_3{^2\!}F_1$ & ${^1\!}F_0{^2\!}F_1+{^1\!}F_1{^2\!}F_0$ & ${^1\!}F_2{^2\!}F_3+{^1\!}F_3{^2\!}F_2$ & ${^1\!}F_0{^2\!}F_3+{^1\!}F_3{^2\!}F_0$ & ${^1\!}F_1{^2\!}F_2+{^1\!}F_2{^2\!}F_1$\\  \hline

Two-photon states & ${|\phi^+\rangle }_{A_1C_1}$ & ${|\psi^+\rangle }_{A_1C_1}$ & ${|\phi^+\rangle }_{B_1C_1}$ & ${|\psi^+\rangle }_{B_1C_1}$ & ${|\phi^+\rangle }_{A_1B_1}$ & ${|\psi^+\rangle }_{A_1B_1}$\\  \hline
\end{tabular}}

\end{table*}

In above, we discussed the success case of entanglement purification of the identity combinations ${|\Phi_i^+\rangle }_1\otimes{|\Phi_i^+\rangle }_2$ from different ensembles. Now we consider the purification of the remaining 12 cross combinations ${|\Phi_i^+\rangle }_1\otimes{|\Phi_j^+\rangle }_2$ $(i\!\ne\! j\!\in\!\left\{0,1,2,3\right\})$, which is discarded in previous MEPPs \cite{earlymepp1,earlymepp2}. 

Similar to Ref. \cite{36-pra-deng-mepp}, if the measurement results of QND are different, we can divide them into three groups: eoe/oeo, oee/eoo and eeo/ooe. Then, three parties can extract high-fidelity two-photon entangled states from the 12 cross combinations of two three-photon systems, as shown in Table \ref{table1}. The following is a brief description of the case where the parity-check results are even, odd, even, or odd, even, odd, after QNDs.

There are only four possible cross combinations when the result is eoe/oeo, that is, $|\Phi_0^+\rangle \otimes|\Phi_2^+\rangle $, $|\Phi_2^+\rangle \otimes|\Phi_0^+\rangle $, $|\Phi_1^+\rangle \otimes|\Phi_3^+\rangle $ and $|\Phi_3^+\rangle \otimes|\Phi_1^+\rangle $. We can rewrite these terms according to the outcomes of the parity-check measurements. Specifically, if the result is oeo, the six-photon system is in the mixed state which is composed of the four states
\begin{subequations}\label{9ad}
\begin{align}
{|\Omega\rangle }_1=&\frac{1}{\sqrt{2}}({|HHH\rangle }_{A_1B_1C_1}{|VHV\rangle }_{A_2B_2C_2}\nonumber\\
&+{|VVV\rangle }_{A_1B_1C_1}{|HVH\rangle }_{A_2B_2C_2}),\\
{|\Omega\rangle }_2=&\frac{1}{\sqrt{2}}({|VHV\rangle }_{A_1B_1C_1}{|HHH\rangle }_{A_2B_2C_2}\nonumber\\
&+{|HVH\rangle }_{A_1B_1C_1}{|VVV\rangle }_{A_2B_2C_2}),\\
{|\Omega\rangle }_3=&\frac{1}{\sqrt{2}}({|VHH\rangle }_{A_1B_1C_1}{|HHV\rangle }_{A_2B_2C_2}\nonumber\\
&+{|HVV\rangle }_{A_1B_1C_1}{|VVH\rangle }_{A_2B_2C_2}),\\
{|\Omega\rangle }_4=&\frac{1}{\sqrt{2}}({|HHV\rangle }_{A_1B_1C_1}{|VHH\rangle }_{A_2B_2C_2}\nonumber\\
&+{|VVH\rangle }_{A_1B_1C_1}{|HVV\rangle }_{A_2B_2C_2}),
\end{align}
\end{subequations}
with the probability of $\frac{1}{2}{^1\!}F_0{^2\!}F_2$, $\frac{1}{2}{^1\!}F_2{^2\!}F_0$, $\frac{1}{2}{^1\!}F_1{^2\!}F_3$ and $\frac{1}{2}{^1\!}F_3{^2\!}F_1$, respectively. In addition, for the case of eoe, the six-photon system is in the mixed state composed of the four states
\begin{subequations}\label{10ad}
\begin{align}
{|\Omega\rangle }_5=&\frac{1}{\sqrt{2}}({|HHH\rangle }_{A_1B_1C_1}{|HVH\rangle }_{A_2B_2C_2}\nonumber\\
&+{|VVV\rangle }_{A_1B_1C_1}{|VHV\rangle }_{A_2B_2C_2}),\\
{|\Omega\rangle }_6=&\frac{1}{\sqrt{2}}({|HVH\rangle }_{A_1B_1C_1}{|HHH\rangle }_{A_2B_2C_2}\nonumber\\
&+{|VHV\rangle }_{A_1B_1C_1}{|VVV\rangle }_{A_2B_2C_2}),\\
{|\Omega\rangle }_7=&\frac{1}{\sqrt{2}}({|VHH\rangle }_{A_1B_1C_1}{|VVH\rangle }_{A_2B_2C_2}\nonumber\\
&+{|HVV\rangle }_{A_1B_1C_1}{|HHV\rangle }_{A_2B_2C_2}),\\
{|\Omega\rangle }_8=&\frac{1}{\sqrt{2}}({|VVH\rangle }_{A_1B_1C_1}{|VHH\rangle }_{A_2B_2C_2}\nonumber\\
&+{|HHV\rangle }_{A_1B_1C_1}{|HVV\rangle }_{A_2B_2C_2}),
\end{align}
\end{subequations}
with the probability of $\frac{1}{2}{^1\!}F_0{^2\!}F_2$, $\frac{1}{2}{^1\!}F_2{^2\!}F_0$, $\frac{1}{2}{^1\!}F_1{^2\!}F_3$ and $\frac{1}{2}{^1\!}F_3{^2\!}F_1$, respectively. 
Actually, the parties may reserve different two-photon systems according to different values of these probabilities, distinct from the same ensemble hypothesis in Ref. \cite{36-pra-deng-mepp}. Generally speaking, coefficients  ${^1\!}F_0$ and ${^2\!}F_0$ are usually much larger compared with other ${^\alpha\!}F_i$. The parties choose to obtain entangled pair $A_1C_1$ in order to obtain high-fidelity two-photon entangled subsystems, by measuring $A_2C_2B_1B_2$ with the basis of $X\!=\!\left \{|\pm\rangle \right \}$ respectively.

Specifically, if the number of the outcomes $|-\rangle $ by measuring $A_2C_2B_1B_2$ is even, then $A_1C_1$ is in the state ${|\phi^+\rangle }_{A_1C_1}\!=\!\frac{1}{\sqrt{2}}{(|HH\rangle \!+\!|VV\rangle )}_{A_1C_1}$ or ${|\psi^+\rangle }_{A_1C_1}\!=\!\frac{1}{\sqrt{2}}{(|HV\rangle \!+\!|VH\rangle )}_{A_1C_1}$, namely, Alice and Charlie obtain the two-photon entangled state ${|\phi^+\rangle }_{A_1C_1}$ or ${|\psi^+\rangle }_{A_1C_1}$ from the six-photon state. If the number of the outcomes $|-\rangle $ is odd, $A_1C_1$ is in the state ${|\phi^-\rangle }_{A_1C_1}\!=\!\frac{1}{\sqrt{2}}{(|HH\rangle \!-\!|VV\rangle )}_{A_1C_1}$ or ${|\psi^-\rangle }_{A_1C_1}\!=\!\frac{1}{\sqrt{2}}{(|HV\rangle\!-\!|VH\rangle )}_{A_1C_1}$, then Alice and Charlie can transform the state ${|\phi^-\rangle }_{A_1C_1}$ into the state ${|\phi^+\rangle }_{A_1C_1}$ or ${|\psi^-\rangle }_{A_1C_1}$ into ${|\psi^+\rangle }_{A_1C_1}$ by performing a phase-flip operation $\sigma_z$ on either photon $A_1$ or photon $C_1$. Thus, they can obtain the two-photon entangled state ${|\phi^+\rangle }_{A_1C_1}$ from the cross combinations $|\Phi_0^+\rangle \otimes|\Phi_2^+\rangle $ and $|\Phi_2^+\rangle \otimes|\Phi_0^+\rangle $ with the probability of ${^1\!}F_0{^2\!}F_2\!+\!{^1\!}F_2{^2\!}F_0$, and obtain ${|\psi^+\rangle }_{A_1C_1}$ from $|\Phi_1^+\rangle \otimes|\Phi_3^+\rangle $ and $|\Phi_3^+\rangle \otimes|\Phi_1^+\rangle $ with the probability of ${^1\!}F_1{^2\!}F_3\!+\!{^1\!}F_3{^2\!}F_1$. A detailed result is shown in Table \ref{table1} for other states of two-photon system obtained from cross combinations after QNDs. Finally, the  two-photon entangled state successfully extracted by two of the three parties can be described in the following:
\begin{align}\label{eq11-13}
{\rho}_{AB}=&({^1\!}F_0{^2\!}F_3+{^1\!}F_3{^2\!}F_0){|\phi^+\rangle }_{AB}\langle \phi^+|\nonumber\\
&+({^1\!}F_1{^2\!}F_2+{^1\!}F_2{^2\!}F_1){|\psi^+\rangle }_{AB}\langle \psi^+|,\\
{\rho}_{AC}=&({^1}F_0{^2}F_2+{^1}F_2{^2}F_0){|\phi^+\rangle }_{AC}\langle \phi^+|\nonumber\\
&+({^1\!}F_1{^2\!}F_3+{^1\!}F_3{^2\!}F_1){|\psi^+\rangle }_{AC}\langle \psi^+|,\\
{\rho}_{BC}=&({^1\!}F_0{^2\!}F_1+{^1\!}F_1{^2\!}F_0){|\phi^+\rangle }_{BC}\langle \phi^+|\nonumber\\
&+({^1\!}F_2{^2\!}F_3+{^1\!}F_3{^2\!}F_2){|\psi^+\rangle }_{BC}\langle \psi^+|.
\end{align}
Suppose that the initial mixed state is symmetric, then the fidelity of three two-photon system ${\rho}_{AB}$ , ${\rho}_{AC}$ and ${\rho}_{BC}$ are equal and they can be simplified as (normalized)
\begin{align}\label{eq14}
\rho^b=F_0^b{|\phi^+\rangle }\langle \phi^+|+F_1^b{|\psi^+\rangle }\langle \psi^+|,
\end{align}
where
\begin{align}\label{eq15}
F_0^b&=\frac{{^1\!}F_0{^2\!}F_1+{^1\!}F_1{^2\!}F_0}{{^1\!}F_0{^2\!}F_1+{^1\!}F_1{^2\!}F_0+2{^1\!}F_1{^2\!}F_1}\nonumber\\
&=\frac{3({^1\!}F_0+{^2\!}F_0-2{^1\!}F_0{^2\!}F_0)}{{^1\!}F_0+{^2\!}F_0-4{^1\!}F_0{^2\!}F_0+2}.
\end{align}

\begin{figure}[ht]
\centering
\includegraphics[width=.45\textwidth]{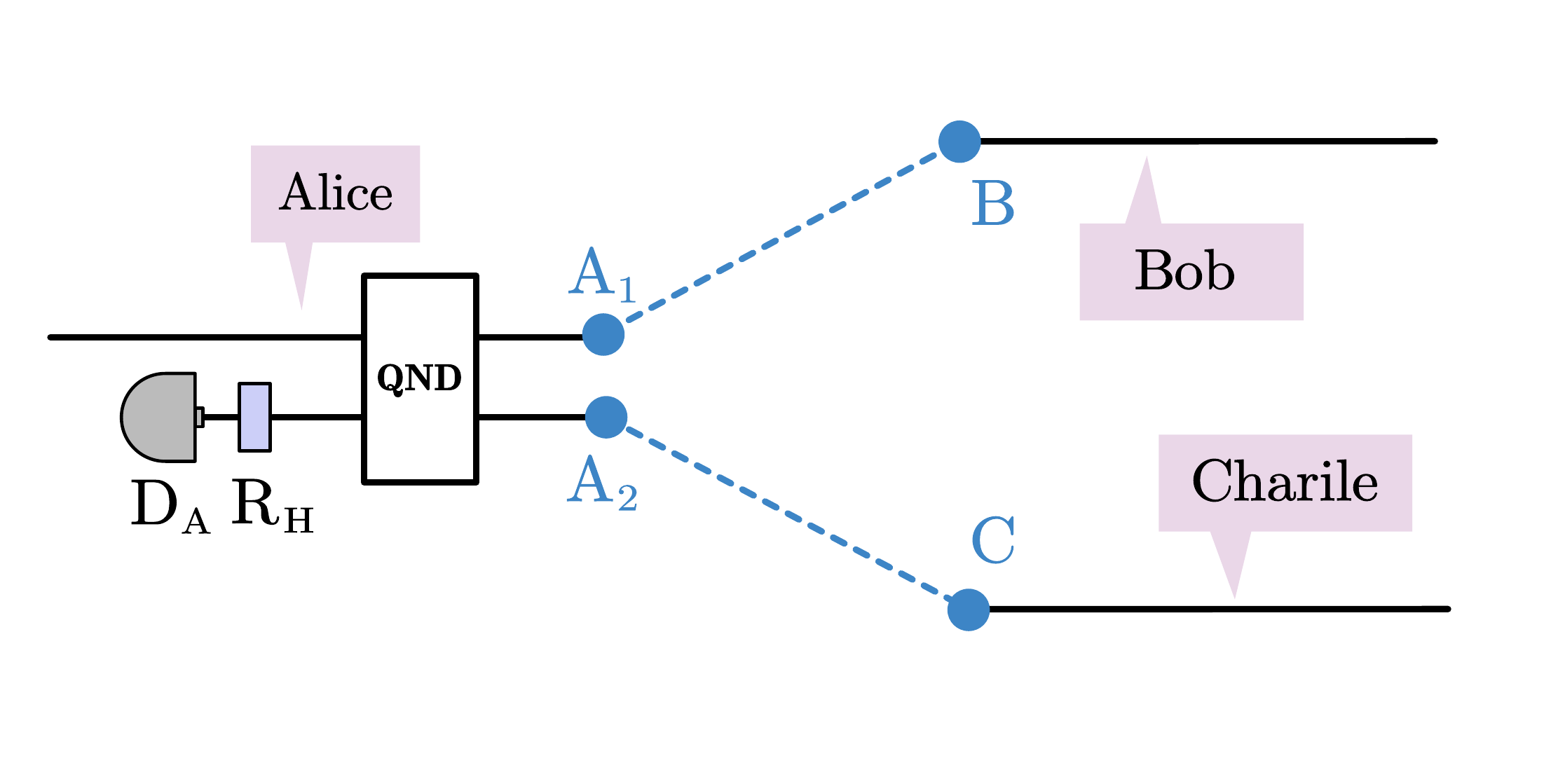}
\caption{The generation of a high-fidelity three-photon entangled system from two two-photon entangled systems with the entanglement link from subspaces \cite{36-pra-deng-mepp}.}
\label{pic2}
\end{figure}
Then, the parties select two pairs from the three two-photon entangled systems, and regenerate three-photon entangled state with the entanglement link from subspaces, like Ref. \cite{36-pra-deng-mepp}. In fact, the general case where the initial entangled ensembles are asymmetric will be very complex, and it's necessary to consider various combinations in multiple rounds of purification and the rational allocation of entangled pairs. Here, we will only take the case of symmetric entangled systems as an example to discuss in detail. 
We selected four photons (marked as $A_1,B,A_2,C$) from the system ${\rho}_{AB}$ and ${\rho}_{AC}$, which is in the state ${\rho}_{A_1B}\otimes{\rho}_{A_2C}$, namely, in the mixture of the four pure states ${|\phi^+\rangle}_{A_1B}\otimes{|\phi^+\rangle }_{A_2C}$, ${|\phi^+\rangle }_{A_1B}\otimes{|\psi^+\rangle }_{A_2C}$, ${|\psi^+\rangle }_{A_1B}\otimes{|\phi^+\rangle }_{A_2C}$ and ${|\psi^+\rangle }_{A_1B}\otimes{|\psi^+\rangle }_{A_2C}$, with the probabilities $F^b_0F^b_0$, $F^b_0F^b_1$, $F^b_1F^b_0$ and $F^b_1F^b_1$, respectively.

As shown in Fig. \ref{pic2}, after Alice performs QND, the results are divided into two groups, one is that the parity-check result is even, and the other is odd. If Alice obtains an even parity, the four-photon system is in the states 
$\frac{1}{\sqrt{2}}{(|HHHH\rangle +|VVVV\rangle )}_{A_1BA_2C}$, $\frac{1}{\sqrt{2}}{(\!|\!H\!H\!H\!V\!\rangle\! +\!|\!V\!V\!V\!H\rangle\! )}_{\!A_1\!B\!A_2C}$, $\frac{1}{\sqrt{2}}(\!|\!H\!V\!H\!H\!\rangle\! +\!
{|\!V\!H\!V\!V\!\rangle \!)}_{\!A_1\!B\!A_2C}$ and $\frac{1}{\sqrt{2}}{(|HVHV\rangle\! +\!|VHVH\rangle )}_{A_1BA_2C}$, with the probabilities $\frac{1}{2}F^b_0F^b_0$, $\frac{1}{2}F^b_0F^b_1$, $\frac{1}{2}F^b_1F^b_0$ and $\frac{1}{2}F^b_1F^b_1$, respectively. Similar to above description, Alice can measure $A_2$ (or $A_1$) with the basis $X\!=\!\left \{|\pm\rangle \right \}$, and then the parties will obtain a three-photon entangled state  $|\Phi_0^+\rangle $, $|\Phi_3^+\rangle $, $|\Phi_2^+\rangle $ and $|\Phi_1^+\rangle $ respectively, with (the result of measurement is $|-\rangle $) 
or without (the result of measurement is $|+\rangle $) a phase-flip operation $\sigma_z$  is performed on the photon $A_1$ (or $A_2$).

If Alice obtains an odd parity , the four-photon entangled state is
$\frac{1}{\sqrt{2}}{(|HHVV\rangle +|VVHH\rangle)}_{A_1BA_2C}$, $\frac{1}{\sqrt{2}}{(\!|\!H\!H\!V\!H\!\rangle \!+\!|\!V\!V\!H\!V\!\rangle\! )}_{\!A_1\!B\!A_2C}$,$\frac{1}{\sqrt{2}}{(\!|\!H\!V\!V\!V\!\rangle\!+\!|\!V\!H\!H\!H\!\rangle\! )}_{\!A_1\!B\!A_2C}$ 
and $\frac{1}{\sqrt{2}}{(|HVVH\rangle \!+\!|VHHV\rangle )}_{A_1BA_2C}$,  with the probabilities $\frac{1}{2}F^b_0F^b_0$, $\frac{1}{2}F^b_0F^b_1$, $\frac{1}{2}F^b_1F^b_0$ and $\frac{1}{2}F^b_1F^b_1$, respectively. The parties can  perform a bit-flip operation $\sigma_x$ on the photons $A_2$ and $C$ (or $A_1$ and $C$) to obtain the same outcomes as the case where the result of QND is even. In other words, they can obtain the following new ensemble for three-photon entangled state using the entanglement link:
\begin{align}\label{eq16}
\rho^{t}=&F^{t}_0|\Phi_0^+\rangle \langle \Phi_0^+|+F_1^{t}|\Phi_1^+\rangle \langle \Phi_1^+|\nonumber\\
&+F_2^{t}|\Phi_2^+\rangle \langle \Phi_2^+|+F_3^{t}|\Phi_3^+\rangle \langle \Phi_3^+|,
\end{align}
with
\begin{align}\label{eq17}
F_0^{t}&=\frac{{F_0^b}^2}{{(F_0^b+F_1^b)}^2}
={F_0^b}^2\nonumber\\
&=\frac{9{({^1\!}F_0+{^2\!}F_0-2{^1\!}F_0{^2\!}F_0)}^2}{{({^1\!}F_0+{^2\!}F_0-4{^1\!}F_0{^2\!}F_0+2)}^2},
\end{align}
and
\begin{align}\label{eq18}
F_1^{t}&=\frac{{F_1^b}^2}{{(F_0^b+F_1^b)}^2}={F_1^b}^2\nonumber\\
F_2^{t}&=F_3^{t}=F_0^bF_1^b.
\end{align}

\begin{figure}[ht]
\centering
\includegraphics[width=0.4\textwidth]{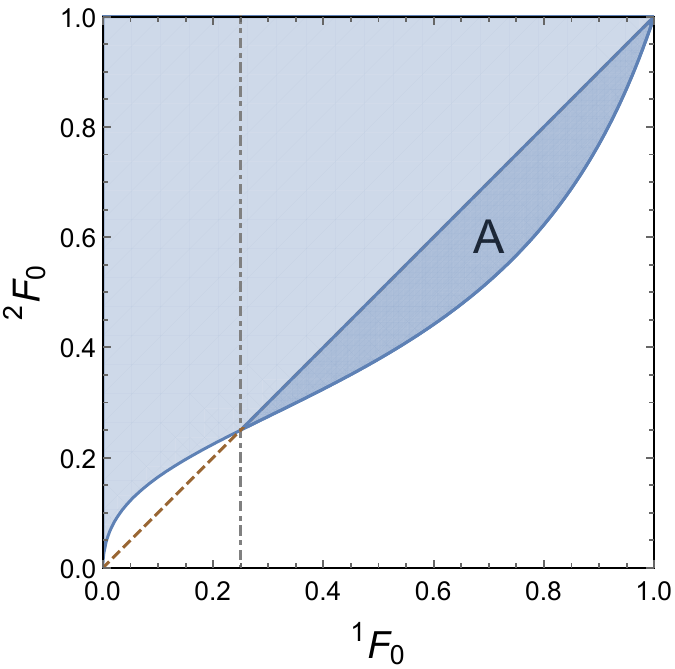}
\caption{The condition for improving the fidelity of new systems generated by entanglement links $F_0^{t}\!>\!{^1}F_0$. The double shaded region A is the  region which satisfy the condition ${^1}F_0\!> \!{^2}F_0$.}
\label{pic3}
\end{figure}

\begin{figure*}[htp]
\centering
\includegraphics[width=0.9\textwidth]{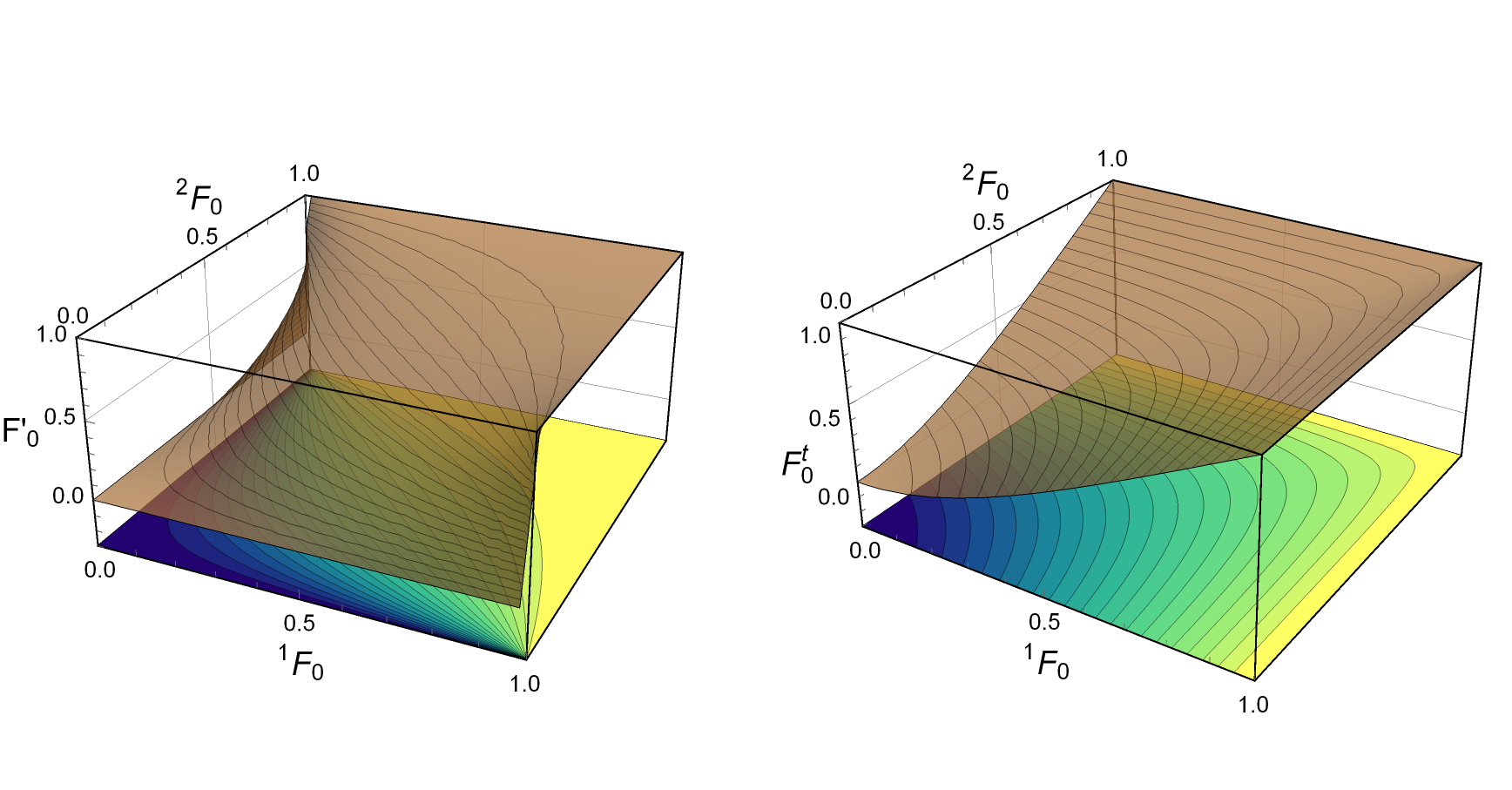}
\caption{The fidelity $F'_0$ and $F_0^{t}$ of the new systems after the correction for bit-flip errors.}
\label{pic4}
\end{figure*}

We consider the condition for successful purification of the cross combinations of the three-photon entangled state transmitted over symmetric noisy channels with the entanglement links, that is, $F_0^{t}\!> \!{^1\!}F_0$. Compared with the case of the same ensembles ${^1\!}F_0\!=\!{^2\!}F_0$ in Ref. \cite{36-pra-deng-mepp}, it's obvious that only the shaded part shown in Fig. \ref{pic3}, is the feasible region where $F_0^{t}\!>\! {^1\!}F_0$ can be established. When ${^1\!}F_0\!=\!{^2\!}F_0$, the feasible region will degenerate into ${^1\!}F_0\!=\!{^2\!}F_0\!>\! 1/4$, namely, the conclusion in Ref. \cite{36-pra-deng-mepp}. However, in the general case of ${^1\!}F_0\!\ne\!{^2\!}F_0$, the feasible region of $F_0^{t}\!>\!{^1\!}F_0$ is very little, which means the parties can only purify the cross combinations successfully in a very small range. Actually, $F_0^{t} > 1/4$ means that the remained mixed state in Eq. \eqref{eq16} can be re-purified in the next round. According to Eq. $\eqref{eq6}$ and $\eqref{eq17}$, we also drawn $F'_0$ and $F_0^{t}$ altered with the initial fidelity $^{1\!}F_0$ and $^{2\!}F_0$ in Fig. \ref{pic4}.

\subsection{MEPP-P1$^\prime$: Direct purification of residual entanglement for bit-flip errors}\label{P1'}
The purification of residual entanglement for two-particle Bell state has been discussed in Ref.\cite{13-oe-diffepp}.
It show that if the initial mixed states are different, the remained states may have entanglement in the case of unsuccessful purification.  In this section, we will show that for the purification of GHZ state, if the fidelity of initial mixed states are different, these cross combinations still exist the residual entanglement which can be reused directly. Here we call such purification scheme as  P1$^\prime$, and scheme P1$^\prime$ is completely consistent with P1 for the treatment of the identity combinations ${|\Phi_i^+\rangle }_1\otimes{|\Phi_i^+\rangle }_2 $, except for the treatment of 12 cross combinations ${|\Phi_i^+\rangle }_1\otimes{|\Phi_j^+\rangle }_2$($i\ne j$). In the following, we will investigate such cross combinations.

We take the situation of oeo and eoe in the measurement results as an example to elaborate in detail. In this case, according to Eq. \eqref{9ad}, the six-photon system $A_1B_1C_1A_2B_2C_2$ is in the states ${|\Omega\rangle} _1$, ${|\Omega\rangle }_2$, ${|\Omega\rangle} _3$ and ${|\Omega\rangle} _4$ with a probability of $\frac{1}{2}{^1\!}F_0{^2\!}F_2$, $\frac{1}{2}{^1\!}F_2{^2\!}F_0$, $\frac{1}{2}{^1\!}F_1{^2\!}F_3$ and $\frac{1}{2}{^1\!}F_3{^2\!}F_1$, respectively. Then they measure the photons $A_2B_2C_2$ of the six-photon system with the basis $X\!=\!\left \{|\pm\rangle \right \}$, and can eventually obtain the quantum system $A_1B_1C_1$ in a new mixed state which is composed of the four states ${|\Phi_0^+\rangle }_{ABC}$, ${|\Phi_2^+\rangle }_{ABC}$, ${|\Phi_1^+\rangle }_{ABC}$ and ${|\Phi_3^+\rangle }_{ABC}$ with the probabilities $\frac{1}{2}{^1\!}F_0{^2\!}F_2$, $\frac{1}{2}{^1\!}F_2{^2\!}F_0$, $\frac{1}{2}{^1\!}F_1{^2\!}F_3$ and $\frac{1}{2}{^1\!}F_3{^2\!}F_1$, respectively. In another case, if the  the measurement results is eoe, the system $A_1B_1C_1A_2B_2C_2$ is in the states ${|\Omega\rangle} _5$, ${|\Omega\rangle }_6$, ${|\Omega\rangle} _7$ and ${|\Omega\rangle} _8$ with the same probabilities as above, by Eq. \eqref{10ad}. They can also obtain the states ${|\Phi_0^+\rangle }_{ABC}$, ${|\Phi_2^+\rangle }_{ABC}$, ${|\Phi_1^+\rangle }_{ABC}$ and ${|\Phi_3^+\rangle }_{ABC}$ with the probabilities $\frac{1}{2}{^1\!}F_0{^2\!}F_2$, $\frac{1}{2}{^1\!}F_2{^2\!}F_0$, $\frac{1}{2}{^1\!}F_1{^2\!}F_3$ and $\frac{1}{2}{^1\!}F_3{^2\!}F_1$, respectively, through the measurement on the photons $A_2B_2C_2$ with the basis $X\!=\!\left \{|\pm\rangle \right \}$.

Therefore, in the cases of eoe and oeo, the parties can obtain a new mixed state composed of $|\Phi_0^+\rangle $, $|\Phi_2^+\rangle $, $|\Phi_1^+\rangle $ and $|\Phi_3^+\rangle $ with the probabilities ${^1\!}F_0{^2\!}F_2$, ${^1\!}F_2{^2\!}F_0$, ${^1\!}F_1{^2\!}F_3$ and ${^1\!}F_3{^2\!}F_1$, respectively, that is (not normalized),
\begin{align}\label{eq19}
\rho''_{ABC(eoe/oeo)}=&{^1\!}F_0{^2\!}F_2|\Phi_0^+\rangle \langle \Phi_0^+|+{^1\!}F_1{^2\!}F_3|\Phi_1^+\rangle \langle \Phi_1^+|\nonumber\\
&+{^1\!}F_2{^2\!}F_0|\Phi_2^+\rangle \langle \Phi_2^+|+{^1\!}F_3{^2\!}F_1|\Phi_3^+\rangle \langle \Phi_3^+|.
\end{align}
Thus, the fidelity of the new system $F''_{0(eoe/oeo)}$ is
\begin{align}\label{eq20}
F''_{0(eoe/oeo)}=\frac{{^1\!}F_0{^2\!}F_2}{{^1\!}F_0{^2\!}F_2+{^1\!}F_2{^2\!}F_0+{^1\!}F_1{^2\!}F_3+{^1\!}F_3{^2\!}F_1}.
\end{align}

Similarly,  if the measurement results are  oee/eoo, and eeo/ooe, they can respectively obtain the new systems (not normalized)
\begin{align}
\rho''_{ABC(oee/eoo)}=&{^1\!}F_0{^2\!}F_1|\Phi_0^+\rangle \langle \Phi_0^+|+{^1\!}F_1{^2\!}F_0|\Phi_1^+\rangle \langle \Phi_1^+|\nonumber\\
&+{^1\!}F_2{^2\!}F_3|\Phi_2^+\rangle \langle \Phi_2^+|+{^1\!}F_3{^2\!}F_2|\Phi_3^+\rangle \langle \Phi_3^+|,\label{eq21}\\
\rho''_{ABC(eeo/ooe)}=&{^1\!}F_0{^2\!}F_3|\Phi_0^+\rangle \langle \Phi_0^+|+{^1\!}F_1{^2\!}F_2|\Phi_1^+\rangle \langle \Phi_1^+|\nonumber\\
&+{^1\!}F_2{^2\!}F_1|\Phi_2^+\rangle \langle \Phi_2^+|+{^1\!}F_3{^2\!}F_0|\Phi_3^+\rangle \langle \Phi_3^+|.\label{eq22}
\end{align}
The fidelity $F''_{0(oee/eoo)}$ and $F''_{0(eeo/ooe)}$ can be written as
\begin{align}\label{eq23}
F''_{0(oee/eoo)}=\frac{{^1\!}F_0{^2\!}F_1}{{^1\!}F_0{^2\!}F_1+{^1\!}F_1{^2\!}F_0+{^1\!}F_2{^2\!}F_3+{^1\!}F_3{^2\!}F_2},
\end{align}
\begin{align}\label{eq24}
F''_{0(eeo/ooe)}=\frac{{^1\!}F_0{^2\!}F_3}{{^1\!}F_0{^2\!}F_3+{^1\!}F_3{^2\!}F_0+{^1\!}F_1{^2\!}F_2+{^1\!}F_2{^2\!}F_1},
\end{align}
respectively.

It should be pointed out that before conducting the second round of entanglement purification, the parties should reverse the diagonal elements of the density matrix in Eq. \eqref{eq19}, \eqref{eq21} and $\eqref{eq22}$ by performing a bit-flip operation $\sigma_x$ on some qubits, and let the probability of the system in the state $|\Phi_0^+\rangle $ is the maximum value. Taking $\rho''_{(eoe/oeo)}$ for example, ABC should choose the largest of the four items ${^1\!}F_0{^2\!}F_2,{^1\!}F_2{^2\!}F_0,{^1\!}F_1{^2\!}F_3,{^1\!}F_3{^2\!}F_1$, as the coefficient of $|\Phi_0^+\rangle \langle \Phi_0^+|$, so that $F''_{0(eoe/oeo)}$ will take the maximum of the four possibilities, namely, $\max\{{^1\!}F_0{^2\!}F_2,{^1\!}F_2{^2\!}F_0,{^1\!}F_1{^2\!}F_3,{^1\!}F_3{^2\!}F_1\}/({^1\!}F_0{^2\!}F_2\!+\!{^1\!}F_2{^2\!}F_0$
$+\!{^1\!}F_1{^2\!}F_3\!+\!{^1\!}F_3{^2\!}F_1)$, and $F'''_0$ obtained from the second round of purification will be larger. This is different from the case of two-photon systems, although we have chosen ${^1\!}F_0\!>\! {^2\!}F_0$, which does not guarantee the state $|\Phi_0^+\rangle $ in system $\rho''$ with the maximum probability. Specifically, if ${^1\!}F_2{^2\!}F_0> {^1\!}F_0{^2\!}F_2$, then Bob should perform a bit-flip operation $\sigma_ X$ on the qubit $B$, so that $|\Phi_0^+\rangle \longleftrightarrow|\Phi_2^+\rangle $ and $|\Phi_1^+\rangle \longleftrightarrow|\Phi_3^+\rangle $.

Supposing $\max \{{^1\!}F_0{^2\!}F_2,\!{^1\!}F_2{^2\!}F_0,\!{^1\!}F_1{^2\!}F_3,\!{^1\!}F_3{^2\!}F_1\}\!=\!{^1\!}F_0{^2\!}F_2$, then $F''_{0(eoe/oeo)}$ can be described by Eq. \eqref{eq20} without any bit-flip operation, and the new fidelity obtained after the second round of purification is
\begin{align}\label{eq25}
F'''_{0(eoe/oeo)}\!=\!\frac{{({^1\!}F_0{^2\!}F_2)}^2}{{({^1\!}F_0{^2\!}F_2)}^2\!+\!{({^1\!}F_2{^2\!}F_0)}^2\!+\!{({^1\!}F_1{^2\!}F_3)}^2\!+\!{({^1\!}F_3{^2\!}F_1)}^2}.
\end{align}
Besides, here we only consider the new system $\rho'''$ from the identity combinations after the second round of purification. In fact, the cross combinations in the second round of purification may still have residual entanglement, which can be purified again.

\begin{figure}[ht]
\centering
\includegraphics[width=0.45\textwidth]{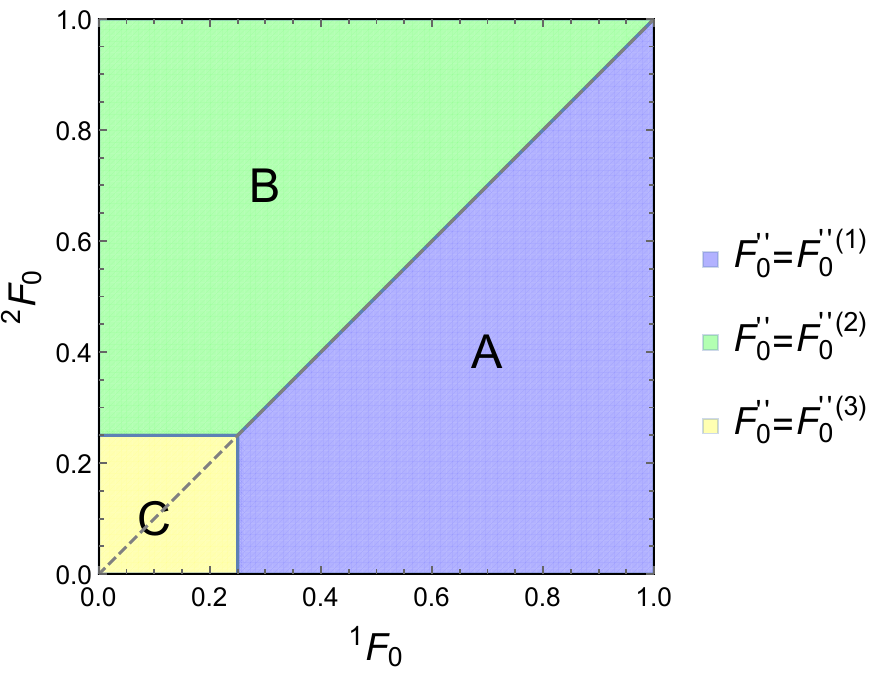}
\caption{The rational choice for $F''_0$ when the three-photon entangled systems are symmetric.}
\label{pic5}
\end{figure}

For the case of symmetric initial entangled ensembles, Eqs. $\eqref{eq20}$ and $\eqref{eq25}$ can be simplified to
\begin{align}\label{eq26}
F''_0&=F''_{0(eoe/oeo)}=F''_{0(oee/eoo)}=F''_{0(eeo/ooe)}\nonumber\\&=\frac{3{^1\!}F_0(1-{^2\!}F_0)}{{^1\!}F_0+{^2\!}F_0-4{^1\!}F_0{^2\!}F_0+2}
\end{align}
and
\begin{align}\label{eq27}
F'''_0&=F'''_{0(eoe/oeo)}=F'''_{0(oee/eoo)}=F'''_{0(eeo/ooe)}\nonumber\\&=\frac{9{{^1\!}F_0}^2{(1\!-\!{^2\!}F_0)}^2}{9{{^1\!}F_0}^2{(1\!-\!{^2\!}F_0)}^2\!+\!9{{^2\!}F_0}^2{(1\!-\!{^1\!}F_0)}^2\!+\!2{(1\!-\!{^1\!}F_0)}^2{(1\!-\!{^2\!}F_0)}^2},
\end{align}
Here we show that in the case of symmetric entangled systems, $\max\{{^1\!}F_0{^2\!}F_2,{^1\!}F_2{^2\!}F_0,{^1\!}F_1{^2\!}F_3,{^1\!}F_3{^2\!}F_1\}$ should be ${^1\!}F_0{^2\!}F_2$, which means the parties don't need to perform any bit-flip operation. There are only three possible choices for $F''_0$, namely,
\begin{align}\label{eq28}
F''^{(1)}_0&=\frac{3{^1\!}F_0(1-{^2\!}F_0)}{{^1\!}F_0+{^2\!}F_0-4{^1\!}F_0{^2\!}F_0+2},\nonumber\\
F''^{(2)}_0&=\frac{3{^2\!}F_0(1-{^1\!}F_0)}{{^1\!}F_0+{^2\!}F_0-4{^1\!}F_0{^2\!}F_0+2},\nonumber\\
F''^{(3)}_0&=\frac{(1-{^1\!}F_0)(1-{^2\!}F_0)}{{^1\!}F_0+{^2\!}F_0-4{^1\!}F_0{^2\!}F_0+2}.
\end{align}
As shown in Fig. \ref{pic5}, the regions of the three cases of $\max\{F''^{(i)}_0\}$ are given. Region A, B, and C represent that $F''_0$ should take values $F''^{(1)}_0$, $F''^{(2)}_0$ and $F''^{(3)}_0$, respectively. Considering the
condition for successfully purifying the identity combinations, i.e. ${^1\!}F_0\!>\! {^2\!}F_0\!>\! 1/4$, it only occurs in region A. Therefore, the rational choice to maximize $F''_0$ is only Eq. \eqref{eq26} if the initial mixed states of  are symmetric.

\begin{figure*}[ht]
\centering
\includegraphics[width=1\textwidth]{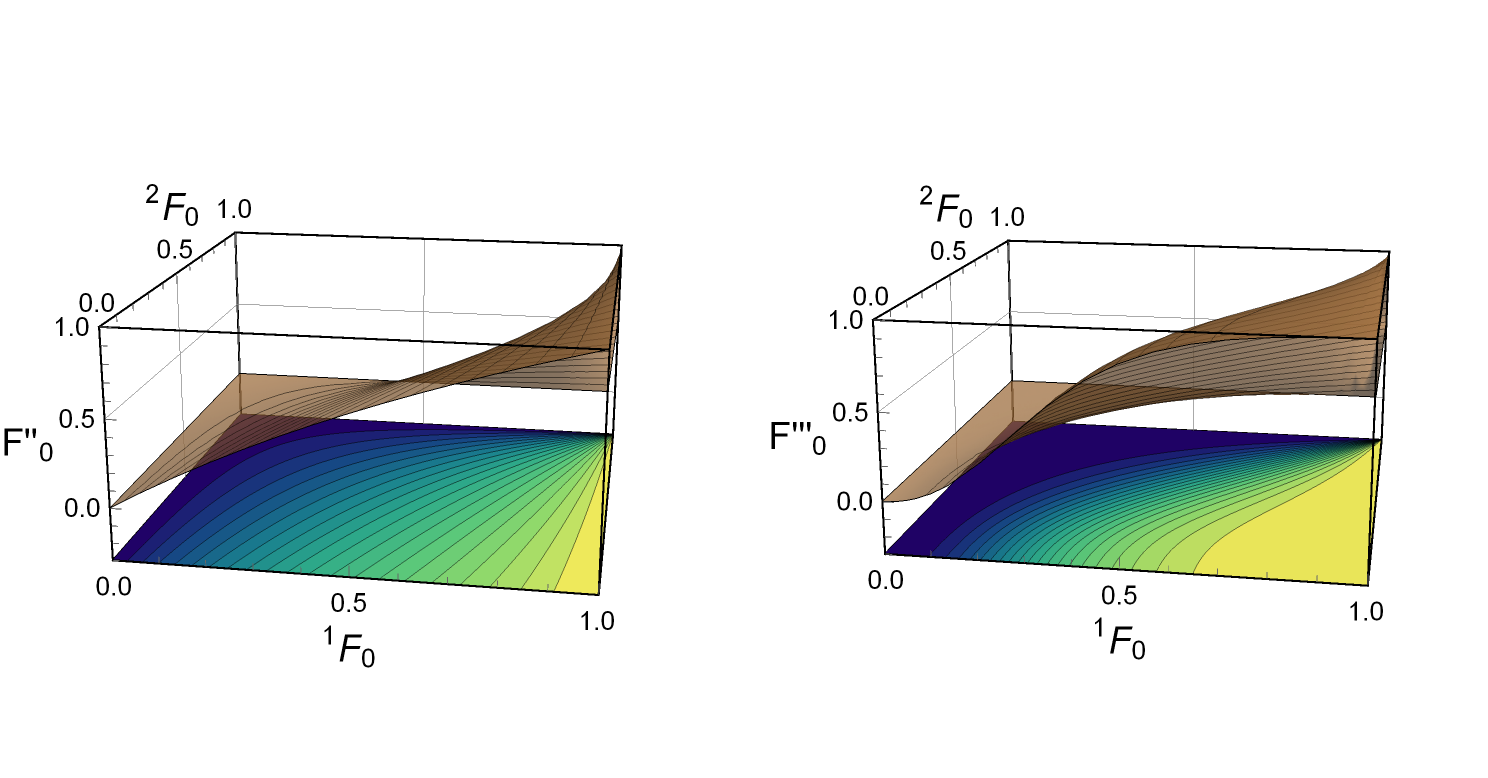}
\caption{The fidelity of the new systems $F''_0$ and $F'''_0$ in scheme P1$^\prime$.}
\label{pic6}
\end{figure*}

The fidelity of the new systems ${F''}$ and ${F'''}$ altered with ${^1}F_0$ and ${^2}F_0$ in the case of symmetric initial ensembles are shown in Fig. \ref{pic6}. One can find the fidelity ${F'''}$ has been significantly improved after the second purification of $\rho''$ obtained from the cross combinations. More detailed analyses are discussed in the Appendix.
In Fig. \ref{pic7}, we show that the shaded area can satisfy the condition $F'''_0> {^1}F_0$, which means the fidelity was successfully improved after the second round of purification. We find that using the cross combinations can successfully improve the fidelity if the initial fidelity ${^1}F_0$ and ${^2}F_0$ are quite different, and overcome the previous problem that the feasible region of scheme P1 is small.

\begin{figure}[ht]
\centering
\includegraphics[width=0.4\textwidth]{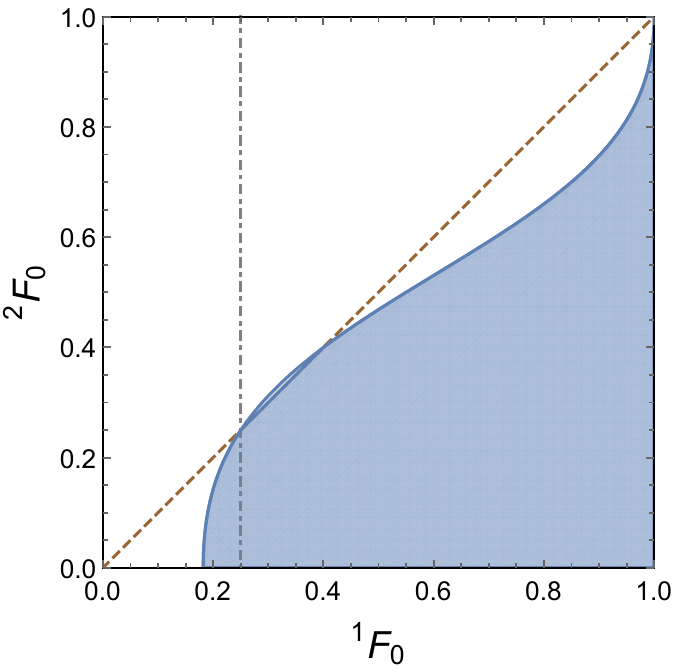}
\caption{The condition for successfully improving the fidelity $F'''_0>{^1}F_0$, after the second round of purification.}
\label{pic7}
\end{figure}

\subsection{N-photon GHZ state entanglement purification for bit-flip errors}\label{N-bit}
The N-photon GHZ states can be written as
\begin{align}\label{eq42}
{|\Phi_{ij...k}^\pm\rangle }_N=\frac{1}{\sqrt{2} } {(|ij...k\rangle \pm|\bar{i}\bar{j}...\bar{k}\rangle )}_N,
\end{align}
where qubit $\bar{i}$ indicates the opposite of $i$, that is, $\bar{i}\!=\!1\!-\!i$, $i,j,...,k\!\in\!\left\{0,1\right\}$, $|0\rangle\! \equiv\!|H\rangle$, $|1\rangle\! \equiv\!|V\rangle $. 
Firstly, according to the above, it's easily to prove that the subsystem in a GHZ state can be extracted from the original system in a GHZ state through the measurement on the other subsystem with the basis $X\!=\!\left \{|\pm\rangle \right \}$. 
Specifically, one can extract the subsystem 1 in the GHZ state 
${|\Phi^+_{ij...m}\rangle }_1=\frac{1}{\sqrt{2} } ({|ij...m\rangle }_1+{|\bar{i}\bar{j}...\bar{m}\rangle}_1)$
from the original system ${|\Phi_{ij...k}^\pm\rangle }_N$ where subsystem 1 and subsystem 2 are entangled. Here system 1 and system 2 are composed of N$_1$ and N$_2$ photons respectively, and N$_1+$N$_2=$N.
To some extent, the idea of MEPP with QNDs is based on this measurement method, that the entangled subsystems can be extracted from various combinations in the case of different parities after QNDs.

Taking N-photon entangled systems with bit-flip errors for example, assume that the GHZ state of N-photon originally prepared is ${|\Phi_0^+\rangle }_N\!=\!\frac{1}{\sqrt{2}}(|HH...H\rangle\!+\!{|VV...V\rangle)}_N$, in addition, the ensemble ${^1}{\rho}$ and  ensemble ${^2}\rho$ of photon systems after the transmission over a noisy channel are respectively in the state
\begin{align}\label{eq43}
{^1\!}{\rho}_{N}=&{^1\!}F_0|\Phi_0^+\rangle \langle \Phi_0^+|+...+{^1\!}F_{ij...k}|\Phi_{ij...k}^+\rangle \langle \Phi_{ij...k}^+|+...\nonumber\\
&+{^1\!}F_{2^{N-1}-1}|\Phi_{2^{N-1}-1}^+\rangle \langle \Phi_{2^{N-1}-1}^+|,\nonumber\\
{^2\!}{\rho}_N=&{^2\!}F_0|\Phi_0^+\rangle \langle \Phi_0^+|+...+{^2\!}F_{ij...k}|\Phi_{ij...k}^+\rangle \langle \Phi_{ij...k}^+|+...\nonumber\\
&+{^2\!}F_{2^{N-1}-1}|\Phi_{2^{N-1}-1}^+\rangle \langle \Phi_{2^{N-1}-1}^+|.
\end{align}
Here ${^\alpha\!}F_{ij...k}$ presents the probability that an N-photon system is in the state ${|\Phi_{ij...k}^+\rangle}_N$, and $ {\textstyle \sum_{m}}{^\alpha\!}F_m\!=\!1$, $m\!\in\!\left\{0,1,...,2^{N-1}\!-\!1\right\}$.

Now we discuss bit-flip error correction for multipartite GHZ state. Similar to Ref.\cite{36-pra-deng-mepp}, if the  parity-check results obtained by all parties  are all odd or all even, they can obtain a new N-photon ensemble which is in the state
\begin{align}\label{eq44}
{\rho}_N'=&F_0'|\Phi_0^+\rangle \langle \Phi_0^+|+...+F_{ij...k}'|\Phi_{ij...k}^+\rangle \langle \Phi_{ij...k}^+|+...\nonumber\\&+F_{2^{N-1}-1}'|\Phi_{2^{N-1}-1}^+\rangle \langle \Phi_{2^{N-1}-1}^+|,
\end{align}
where
\begin{align}\label{eq45}
F'_0\!=\!\frac{{^1\!}F_0{^2\!}F_0}{{^1\!}F_0{^2\!}F_0\!+\!...\!+\!{^1\!}F_{ij...k}{^2\!}F_{ij...k}\!+\!...\!+\!{^1\!}F_{2^{N\!-\!1}\!-\!1}{^2\!}F_{2^{N\!-\!1}\!-\!1}}.
\end{align}
The condition of the fidelity of the new system $F'_0\!>\!{^1}F_0$ is
\begin{align}\label{eq46}
&{^2\!}F_0\!>\!\frac{{^1\!}F_1{^2\!}F_1\!+\!...\!+\!{^1\!}F_{i\!j\!...k}{^2\!}F_{i\!j\!...k}\!+\!...\!+\!{^1\!}F_{2^{N\!-\!1}\!-\!2}{^2\!}F_{2^{N\!-\!1}\!-\!2}}{2\!-\!2{^1\!}F_0\!-\!{^1\!}F_1\!-\!...\!-\!{^1\!}F_{i\!j\!...k}\!-\!...\!-\!{^1\!}F_{2^{N\!-\!1}\!-\!2}}+\nonumber\\
&\frac{(\!1\!\!-\!\!{^1\!}F\!_0\!\!-\!\!...\!\!-\!\!{^1\!}F\!_{\!i\!j\!...k}\!\!-\!\!...\!\!-\!\!{^1\!}F\!_{\!2\!^{N\!-\!1}\!-\!2}\!)(\!1\!\!-\!\!{^2\!}F\!_1\!\!-\!\!...\!\!-\!\!{^2\!}F\!_{\!i\!j\!...k}\!\!-\!\!...\!\!-\!\!{^2\!}F\!_{\!2\!^{N\!-\!1}\!-\!2}\!)}{2\!-\!2{^1\!}F_0\!-\!{^1\!}F_1\!-\!...\!-\!{^1\!}F_{\!i\!j\!...k}\!-\!...\!-\!{^1\!}F_{2^{N\!-\!1}\!-\!2}}.
\end{align}
When the noisy channels are symmetric, the fidelity of the state $|\Phi_0^+\rangle$ will be improved by the conventional MEPP if
\begin{align}\label{eq47}
{^2\!}F_0> \frac{1}{2^{N-1}}.
\end{align}
That is, if ${^1\!}F_0\!>\!{^2\!}F_0\!>\!1/2^{N-1}$ then $F'_0\!>\!{^1\!}F_0$. Meanwhile the lower bound of ${^2\!}F_0$ is $1/2^{N-1}$, consistent with the condition for multipartite entanglement purification in the case where the ensembles are the same and symmetric, and it will revert to this specific case if ${^1\!}F_0\!=\!{^2\!}F_0$.

We next discuss the other cross combinations ${|\Phi_{ij...k}^+\rangle }_1\otimes{|\Phi_{lr...q}^+\rangle }_2$ (where ${i,j,...,k,l,r,...,q}\!\in\!\{0,1\}$ and ${i\!\ne\!l}\bigvee {j\!\ne\!r}\bigvee...\bigvee {k\!\ne\!q}$). Like the above description, there are two approaches P1 and P1$^\prime$ to reuse the cross combinations.

\subsubsection{Approach of entanglement links for N-photon GHZ state }
In this part, we will describe the approach of entanglement link, namely, P1 for N-photon GHZ state with different initial fidelities. Using the approach P1, the parties can obtain some different N$^\prime$-photon ($2\!\le\!N^{\prime}\!<\!N$) subsystems with high-fidelity from the cross combinations according to the different parity-check results, and then regenerate some high-fidelity N-photon entangled systems with entanglement links from subspaces. The more qubits the entangled system contains, the more kinds of combinations and the more complex the quantum circuit will be. We first take the four-photon case for example to briefly explain the general law of extracting N$^\prime$-photon entangled systems from different initial ensembles, according to the parity-check results using QNDs.

The number of entangled photons N$^\prime$ obtained from the four-photon entangled system can be $2$ or $3$. According to the results of QNDs, the parity check results are divided into two groups, the first is eeeo/oooe, eeoe/ooeo, eoee/oeoo, oeee/eooo, and the other is eeoo/ooee, eoeo/oeoe, eooe/oeeo. There are 14 different results (excluding the identity combinations eeee/oooo) in total, corresponding to 56 cross combinations. In the general case of different ensembles, the parties may actually reserve different qubits and even different photon numbers according to different values of ${^\alpha\!}F_i$, distinct from the same ensemble hypothesis in Ref. \cite{36-pra-deng-mepp}. 

Taking the measurement result eeoo for instance, in the case of some specific parameters, they may obtain other possible entangled systems such as two-photon entangled state ${\rho}_{AC}$ or three-photon entangled state ${\rho}_{ABC}$, etc., instead of ${\rho}_{AB}/{\rho}_{CD}$, which can never occur in the case of the same ensembles. In most cases, ${^1\!}F_0$ and ${^2\!}F_0$ are much larger compared with other ${^\alpha\!}F_i$, and there is almost little difference among other ${^\alpha\!}F_i$. Therefore, in order to obtain high-fidelity entangled states, the parties will choose to extract three-photon entangled states and two-photon entangled states respectively from the above two classifications, similar to  Ref. \cite{36-pra-deng-mepp}.

Specifically, in the former case the parties can get some high-fidelity three-photon entangled systems ${\rho}_{ABC}$ (corresponding to eeeo/oooe), ${\rho}_{ABD}$ (eeoe/ooeo), ${\rho}_{ACD}$ (eoee/oeoo) and ${\rho}_{BCD}$ (oeee/eooo) respectively through the operations similar to Sec. \ref{P1}. For ${\rho}_{ABC}$, the density matrix of the new system is (not normalized)
\begin{align}\label{eq48}
{\rho\!}_{ABC}\!=&({^1\!}F\!_0\!{^2\!}F\!_1\!\!+\!\!{^1\!}F\!_1\!{^2\!}F\!_0)|\Phi_0^+\rangle \!\langle \Phi_0^+|\!\!+\!\!({^1\!}F\!_2\!{^2\!}F\!_3\!\!+\!\!{^1\!}F\!_3\!{^2\!}F\!_2)|\Phi_1^+\rangle\! \langle \Phi_1^+|\nonumber\\
&+\!\!({^1\!}F\!_4\!{^2\!}F\!_5\!\!+\!\!{^1\!}F\!_5\!{^2\!}F\!_4)|\Phi_2^+\rangle \!\langle \Phi_2^+|\!\!+\!\!({^1\!}F\!_6\!{^2\!}F\!_7\!\!+\!\!{^1\!}F\!_7\!{^2\!}F\!_6)|\Phi_3^+\rangle\! \langle \Phi_3^+|.
\end{align}
The others can also be given the corresponding formula description similarly. For the latter case, they can get high-fidelity two-photon systems ${\rho}_{AB}/{\rho}_{CD}$ (eeoo/ooee), ${\rho}_{AC}/{\rho}_{BD}$ (eoeo/oeoe) and ${\rho}_{AD}/{\rho}_{BC}$ (eooe/oeeo) respectively through the similar operations. For instance, if the result of QNDs is eeoo/ooee, the parties can obtain two results according to the different qubits they choose to measure. If they measure the photons in $C_1D_1A_2B_2C_2D_2$, then they will obtain ${\rho}_{A_1B_1}$, and if they measure the photons in $A_1B_1A_2B_2C_2D_2$ they can obtain ${\rho}_{C_1D_1}$. The density matrix of the two new systems are respectively (not normalized)
\begin{align}\label{eq49}
{\rho}_{AB}\!=&({^1\!}F_0{^2\!}F_3\!+\!{^1\!}F_3{^2\!}F_0\!+\!{^1\!}F_1{^2\!}F_2\!+\!{^1\!}F_2{^2\!}F_1){|\phi^+\rangle }_{AB}\langle \phi^+|\nonumber\\
&+\!({^1\!}F_4{^2\!}F_7\!+\!{^1\!}F_7{^2\!}F_4\!+\!{^1\!}F_5{^2\!}F_6\!+\!{^1\!}F_6{^2\!}F_5){|\psi^+\rangle }_{AB}\langle \psi^+|,\nonumber\\
{\rho}_{CD}\!=&({^1\!}F_0{^2\!}F_3\!+\!{^1\!}F_3{^2\!}F_0\!+\!{^1\!}F_4{^2\!}F_7\!+\!{^1\!}F_7{^2\!}F_4){|\phi^+\rangle }_{CD}\langle \phi^+|\nonumber\\
&+\!({^1\!}F_1{^2\!}F_2\!+\!{^1\!}F_2{^2\!}F_1\!+\!{^1\!}F_5{^2\!}F_6\!+\!{^1\!}F_6{^2\!}F_5){|\psi^+\rangle }_{CD}\langle \psi^+|.
\end{align}
For these two results, the parties can choose properly in the light of the parameters of the initial entangled ensembles, the allocation of entanglement resources and other factors. 

According to Eq. \eqref{eq48} and \eqref{eq49}, they can regenerate the high-fidelity four-photon systems through the similar operations in Sec. \ref{P1}. It is noteworthy that the analysis of entanglement links will be more complex in the case of the multipartite entanglement purification for N-photon systems, because there are more kinds of combinations. For four-photon systems, the parties can obtain a four-photon entangled systems from the combination of two three-photon subsystems or three two-photon entangled subsystems with entanglement links. Certainly, it can also be completed by a three-photon entangled subsystem and a two-photon entangled subsystem. In addition, each combination may also be distinguished into different schemes according to symmetry, as mentioned in Ref. \cite{36-pra-deng-mepp}. It can be seen that the operation of regenerating high-fidelity N-photon system from multiple N$^\prime$-photon systems through entangled links is similar to Sec. \ref{P1}, whereas its complexity can be much higher and it also needs to allocate the combination properly according to the proportion of the actual entanglement resources.

Extending the analysis of four-photon systems to the case of N-photon, we find that the number of photons in the entangled subsystem can be classified according to the different parity-check results, and the minimum number of photons in the subsystem extracted from the N-photon systems should be $\lfloor \frac{N+1}{2}  \rfloor$. Therefore, the value range of N$^\prime$ is $\left[ \lfloor \frac{N+1}{2}  \rfloor ,N\!\!-\!\!1\right]$.

\begin{table*}
\centering  
\caption{The parity-check results and the corresponding states of four-photon system.}  
\label{table3} 
\resizebox{0.8\linewidth}{!}{
\begin{tabular}{c||c|c|c|c|c|c|c}
& eeeo/oooe & eeoe/ooeo & eeoo/ooee & eoee/oeoo & eoeo/oeoe & eooe/oeeo & eooo/oeee \\ \hline

$|\Phi^+_0\rangle $ & ${^1\!}F_0{^2\!}F_1$ & ${^1\!}F_0{^2\!}F_2$ & ${^1\!}F_0{^2\!}F_3$ & ${^1\!}F_0{^2\!}F_4$ & ${^1\!}F_0{^2\!}F_5$ & ${^1\!}F_0{^2\!}F_6$ & ${^1\!}F_0{^2\!}F_7$ \\  \hline
$|\Phi^+_1\rangle $ & ${^1\!}F_1{^2\!}F_0$ & ${^1\!}F_1{^2\!}F_3$ & ${^1\!}F_1{^2\!}F_2$ & ${^1\!}F_1{^2\!}F_5$ & ${^1\!}F_1{^2\!}F_4$ & ${^1\!}F_1{^2\!}F_7$ & ${^1\!}F_1{^2\!}F_6$ \\  \hline
$|\Phi^+_2\rangle $ & ${^1\!}F_2{^2\!}F_3$ & ${^1\!}F_2{^2\!}F_0$ & ${^1\!}F_2{^2\!}F_1$ & ${^1\!}F_2{^2\!}F_6$ & ${^1\!}F_2{^2\!}F_7$ & ${^1\!}F_2{^2\!}F_4$ & ${^1\!}F_2{^2\!}F_5$ \\  \hline
$|\Phi^+_3\rangle $ & ${^1\!}F_3{^2\!}F_2$ & ${^1\!}F_3{^2\!}F_1$ & ${^1\!}F_3{^2\!}F_0$ & ${^1\!}F_3{^2\!}F_7$ & ${^1\!}F_3{^2\!}F_6$ & ${^1\!}F_3{^2\!}F_5$ & ${^1\!}F_3{^2\!}F_4$ \\  \hline
$|\Phi^+_4\rangle $ & ${^1\!}F_4{^2\!}F_5$ & ${^1\!}F_4{^2\!}F_6$ & ${^1\!}F_4{^2\!}F_7$ & ${^1\!}F_4{^2\!}F_0$ & ${^1\!}F_4{^2\!}F_1$ & ${^1\!}F_4{^2\!}F_2$ & ${^1\!}F_4{^2\!}F_3$ \\  \hline
$|\Phi^+_5\rangle $ & ${^1\!}F_5{^2\!}F_4$ & ${^1\!}F_5{^2\!}F_7$ & ${^1\!}F_5{^2\!}F_6$ & ${^1\!}F_5{^2\!}F_1$ & ${^1\!}F_5{^2\!}F_0$ & ${^1\!}F_5{^2\!}F_3$ & ${^1\!}F_5{^2\!}F_2$ \\  \hline
$|\Phi^+_6\rangle $ & ${^1\!}F_6{^2\!}F_7$ & ${^1\!}F_6{^2\!}F_4$ & ${^1\!}F_6{^2\!}F_5$ & ${^1\!}F_6{^2\!}F_2$ & ${^1\!}F_6{^2\!}F_3$ & ${^1\!}F_6{^2\!}F_0$ & ${^1\!}F_6{^2\!}F_1$ \\  \hline
$|\Phi^+_7\rangle $ & ${^1\!}F_7{^2\!}F_6$ & ${^1\!}F_7{^2\!}F_5$ & ${^1\!}F_7{^2\!}F_4$ & ${^1\!}F_7{^2\!}F_3$ & ${^1\!}F_7{^2\!}F_2$ & ${^1\!}F_7{^2\!}F_1$ & ${^1\!}F_7{^2\!}F_0$ \\  \hline
\end{tabular}}
\end{table*}
\subsubsection{Direct purification of residual entanglement for N-photon GHZ state}
In above subsection, we have described the general process of scheme P1, that the parties can extract high-fidelity N$^\prime$-photon entangled subsystems from N-photon systems, and regenerate high-fidelity N-photon entangled systems using the approach of entanglement link. Next, we will give the description of the general process of scheme P1$^\prime$, i.e., the direct purification of residual entanglement. We still take four-photon systems for example to introduce  scheme P1$^\prime$. After the parity classification and some operations, the results are shown in Table \ref{table3} (without $\sigma_x$ yet). For instance, if the parity-check result is eeeo/oooe, the state of the four-photon system obtained by the parties is (not normalized)
\begin{align}\label{eq50}
\rho''_{ABC\!D(eeeo/oooe)}=&{^1\!}F_0{^2\!}F_1|\Phi_0^+\rangle \langle \Phi_0^+|+{^1\!}F_1{^2\!}F_0|\Phi_1^+\rangle \langle \Phi_1^+|\nonumber\\
&+{^1\!}F_2{^2\!}F_3|\Phi_2^+\rangle \langle \Phi_2^+|+{^1\!}F_3{^2\!}F_2|\Phi_3^+\rangle \langle \Phi_3^+|\nonumber\\
&+{^1\!}F_4{^2\!}F_5|\Phi_4^+\rangle \langle \Phi_4^+|+{^1\!}F_5{^2\!}F_4|\Phi_5^+\rangle \langle \Phi_5^+|\nonumber\\
&+{^1\!}F_6{^2\!}F_7|\Phi_6^+\rangle \langle \Phi_6^+|+{^1\!}F_7{^2\!}F_6|\Phi_7^+\rangle \langle \Phi_7^+|.
\end{align}
In  Table \ref{table3},  obviously, for these 56 cross combinations, a total of 7 different ensembles can be obtained in term of different parities check results. With $\rho'_{ABCD}$ obtained from the identity combinations considered, these 8 new ensembles can be purified independently in multiple rounds. Moreover, for these 7 ensembles obtained from the cross combinations, a bit-flip operation $\sigma_x$ is required for some qubits before next purification, so that the probability that the system is in the state $|\Phi_0^+\rangle $ will be the largest, namely, the coefficient of $|\Phi_0^+\rangle $ should actually take the maximum value in each column of Table \ref{table3}. Then, the parties can obtain high-fidelity four-photon entangled systems by the second round of purification.

It can be found that the four-photon GHZ states have lower symmetry than the three-photon GHZ state in Eq. \eqref{eq1}, and $F''_0$ obtained from the cross combinations will also be more complex, which is the inevitable complexity of multipartite entangled systems. Likewise, in the case of N-photon entangled states, we can obtain a table in the shape of ${(2^{N\!-\!1})}\!\times\! {(2^{N\!-\!1}\!\!-\!\!1)}$, which corresponds ${(2^{N\!-\!1})}\!\times\! {(2^{N\!-\!1}\!\!-\!\!1)}$ types of cross combinations ${|\Phi_{ij...k}^+\rangle }_1\otimes{|\Phi_{lr...q}^+\rangle }_2$ respectively, where ${i,j,...,k,l,r,...,q}\!\in\!\{0,1\}$ and ${i\!\ne\!l}\bigvee {j\!\ne\!r}\bigvee...\bigvee {k\!\ne\!q}$. Each column of the table represents a new N-photon system (each system needs the corresponding $\sigma_x$ operation according to the known parameters of the initial ensembles in advance), then the parties can obtain $2^{N\!-\!1}\!-\!1$ types of N-photon systems ${\rho}''_N$ that can be reused for the multiple rounds of purification. Besides, compared with scheme P1, in scheme P1$^\prime$ the parties do not need to use complex entanglement links to complete the conversion between the systems composed of different photon (qubit) numbers. 
Lastly, for the selection of $F''_0$ in Sec. \ref{P1'}, here we extend it to the general case of symmetric N-photon entangled states, that is, ${^\alpha\!}F_i\!=\!(1\!-\!{^\alpha\!}F_0)/(2^{N\!-\!1}\!-\!1)$, $\alpha\!\in\!\left\{1,2\right\}$, $i\!\in\!\left\{0,1,...,2^{N\!-\!1}\!-\!1\right\}$, $N\!\ge\! 3$, and there are also three possible choices for $F''_0$, namely,
\begin{align}\label{eq51}
F''^{(1)}_0&=\frac{(2^{N-1}\!-\!1){^1\!}F_0(1-{^2\!}F_0)}{{^1\!}F_0+{^2\!}F_0-(2^{N-1}){^1\!}F_0{^2\!}F_0+2^{N-1}-2},\nonumber\\
F''^{(2)}_0&=\frac{(2^{N-1}\!-\!1){^2\!}F_0(1-{^1\!}F_0)}{{^1\!}F_0+{^2\!}F_0-(2^{N-1}){^1\!}F_0{^2\!}F_0+2^{N-1}-2},\nonumber\\
F''^{(3)}_0&=\frac{(1-{^1\!}F_0)(1-{^2\!}F_0)}{{^1\!}F_0+{^2\!}F_0-(2^{N-1}){^1\!}F_0{^2\!}F_0+2^{N-1}-2}.
\end{align}
\textbf{Similar to Fig. \ref{pic5}, we can also give the regions of the three cases of $\max\{F''^{(i)}_0\}$ respectively. Meanwhile considering the assumption ${^1\!}F_0\!>\! {^2\!}F_0$ and the condition for successfully purifying the identity combinations ${^2\!}F_0\!> \!1/2^{N\!-\!1}$, there is the limit of ${^1\!}F_0\!>\! {^2\!}F_0\!>\! 1/2^{N\!-\!1}$, which occurs to be the case where $F''_0$ selects $F''^{(1)}_0$.}

\section{Multipartite entanglement purification for phase-flip errors with non-identical states}\label{phase}

\begin{table*}[t]
	\begin{minipage}{0.49\linewidth}
		\centering	
		\caption{The states obtained from \textbf{the identity combinations} with corresponding probabilities and the unitary operations.}
		\label{table2a}
		\resizebox{1\textwidth}{!}{
\begin{tabular}{c||c||c||c} 
\hline
Parities & States & Probabilities & Unitary operations   \\ \hline

\multirow{2}{*}{eee} & $|\phi\rangle _0$ & $\frac{1}{4}{^1\!}P_0{^2\!}P_0$ & \multirow{2}{*}{$I_{_{A_2}}\!\otimes\! I_{_{B_2}}\!\otimes\! I_{_{C_2}}$}\\ \cline{2-3}
					& $|\phi'\rangle _0$ & $\frac{1}{4}{^1\!}P_1{^2\!}P_1$ \\ \hline

\multirow{2}{*}{eoo} & $|\phi\rangle _1$ & $\frac{1}{4}{^1\!}P_0{^2\!}P_0$ & \multirow{2}{*}{$I_{_{A_2}}\!\otimes\! {\sigma}_{x_{B_2}}\!\otimes\! {\sigma}_{x_{C_2}}$}\\ \cline{2-3}
					& $|\phi'\rangle _1$ & $\frac{1}{4}{^1\!}P_1{^2\!}P_1$ \\ \hline
					
\multirow{2}{*}{oeo} & $|\phi\rangle _2$ & $\frac{1}{4}{^1\!}P_0{^2\!}P_0$ & \multirow{2}{*}{${\sigma}_x{_{A_2}}\!\otimes\! I_{_{B_2}}\!\otimes\! {\sigma}_{x_{C_2}}$}\\ \cline{2-3}
					& $|\phi'\rangle _2$ & $\frac{1}{4}{^1\!}P_1{^2\!}P_1$ \\ \hline
					
\multirow{2}{*}{ooe} & $|\phi\rangle _3$ & $\frac{1}{4}{^1\!}P_0{^2\!}P_0$ & \multirow{2}{*}{${\sigma}_{x_{A_2}}\!\otimes\! {\sigma}_{x_{B_2}}\!\otimes\! I_{_{C_2}}$}\\ \cline{2-3}
					& $|\phi'\rangle _3$ & $\frac{1}{4}{^1\!}P_1{^2\!}P_1$ \\ \hline

\end{tabular}
		}
		
	\end{minipage}
	\hfill
	\begin{minipage}{0.49\linewidth}  
		\centering	
		\caption{The states obtained from \textbf{the cross combinations} with corresponding probabilities and the unitary operations.}
		\label{table2b}
		\resizebox{1\textwidth}{!}{
			\begin{tabular}{c||c||c||c} 
\hline
Parities & States & Probabilities & Unitary operations  \\ \hline

\multirow{2}{*}{ooo} & $|\psi\rangle _0$ & $\frac{1}{4}{^1\!}P_0{^2\!}P_1$ & \multirow{2}{*}{${\sigma}_{x_{A_2}}\!\otimes\! {\sigma}_{x_{B_2}}\!\otimes\! {\sigma}_{x_{C_2}}$}\\ \cline{2-3}
					& $|\psi'\rangle _0$ & $\frac{1}{4}{^1\!}P_1{^2\!}P_0$ \\ \hline

\multirow{2}{*}{oee} & $|\psi\rangle _1$ & $\frac{1}{4}{^1\!}P_0{^2\!}P_1$ & \multirow{2}{*}{${\sigma}_{x_{A_2}}\!\otimes\! I_{_{B_2}}\!\otimes\! I_{_{C_2}}$}\\ \cline{2-3}
					& $|\psi'\rangle _1$ & $\frac{1}{4}{^1\!}P_1{^2\!}P_0$ \\ \hline
					
\multirow{2}{*}{eoe} & $|\psi\rangle _2$ & $\frac{1}{4}{^1\!}P_0{^2\!}P_1$ & \multirow{2}{*}{$I_{_{A_2}}\!\otimes\! {\sigma}_{x_{B_2}}\!\otimes\! I_{_{C_2}}$}\\ \cline{2-3}
					& $|\psi'\rangle _2$ & $\frac{1}{4}{^1\!}P_1{^2\!}P_0$ \\ \hline
					
\multirow{2}{*}{eeo} & $|\psi\rangle _3$ & $\frac{1}{4}{^1\!}P_0{^2\!}P_1$ & \multirow{2}{*}{$I_{_{A_2}}\!\otimes\! I_{_{B_2}}\!\otimes\! {\sigma}_{x_{C_2}}$}\\ \cline{2-3}
					& $|\psi'\rangle _3$ & $\frac{1}{4}{^1\!}P_1{^2\!}P_0$ \\ \hline

\end{tabular}
		}
	\end{minipage}
\end{table*}

In this section, we will discuss the multipartite entanglement purification for phase-flip errors.  We also consider three-photon GHZ states first, and then extend it to the N-photon GHZ states.  Similar to the entanglement purification of two-photon systems, the parties cannot correct phase-flip errors directly, which is different from the correction for bit-flip errors. However, a phase-flip error can be converted into a bit-flip error with a Hadamard operation. For three-photon systems, the states $|\Phi_0^+\rangle $ and $|\Phi_0^-\rangle $ in Eq. \eqref{eq1} can be converted into the states $|\Psi^+\rangle $ and $|\Psi^-\rangle $ respectively as follows:
\begin{align}\label{eq52}
|\Psi^+\rangle &=\frac{1}{2}(|HHH\rangle\! +\!|HVV\rangle \!+\!|VHV\rangle \!+\!|VVH\rangle ),\nonumber\\
|\Psi^-\rangle &=\frac{1}{2}(|HHV\rangle \!+\!|HVH\rangle\! +\!|VHH\rangle\! +\!|VVV\rangle ),
\end{align}
after a Hadamard operation on each photon. 
It can be seen that the conversion between phase-flip errors and bit-flip errors in three-photon GHZ states is more complex than that in Bell states, and we cannot exploit simply Hadamard operations to complete the conversion between a phase-flip error and a single bit-flip error in three-photon GHZ states, which is different from two-photon Bell states with higher symmetry. Therefore, we cannot simply regard the multipartite entanglement purification as a direct extension of the two-particle case, and the parties cannot use the equipment shown in Fig. \ref{pic1} to purify the states in Eq. \eqref{eq52} directly. However, due to the different numbers of the polarization state $|V\rangle $ in these two three-photon states, the relative probability of $|\Psi^-\rangle $ can be reduced according to its parity characteristics to accomplish the purification.

Similarly, in the general case of different ensembles, after performing the Hadamard operations, Alice, Bob and Charlie share the source ensemble ${^1\!}\rho$ and the target ensemble ${^2\!}\rho$ as
\begin{align}\label{eq53}
{^1\!}{\rho}_{ABC}={^1\!}P_0|\Psi^+\rangle \langle \Psi^+|+{^1\!}P_1|\Psi^-\rangle \langle \Psi^-|,\nonumber\\
{^2\!}{\rho}_{ABC}={^2\!}P_0|\Psi^+\rangle \langle \Psi^+|+{^2\!}P_1|\Psi^-\rangle \langle \Psi^-|,
\end{align}
where ${^\alpha\!}P_0\!+\!{^\alpha\!}P_1\!=\!1$, $\alpha\!\in\! \left\{1,2\right\}$, and suppose that ${^1\!}P_0\!>\! {^2\!}P_0$. Likewise, we still need to discuss the identity combinations and the cross combinations separately.

For the identity combinations $|\Psi^+\rangle \otimes|\Psi^+\rangle $ and $|\Psi^-\rangle \otimes|\Psi^-\rangle $, the relationship among the outcomes of the parity-check results measured by Alice, Bob and Charlie, and the states of the quantum system composed of the six photons $A_1B_1C_1A_2B_2C_2$ and their probabilities is shown in Table \ref{table2a}, where 
\begin{align}\label{eq54}
&{|\phi\rangle}_0=\!\frac{1}{2}(|\!H\!H\!H\!H\!H\!H\!\rangle \!\!+\!\!|\!H\!V\!V\!H\!V\!V\!\rangle\!\!+\!\!|\!V\!H\!V\!V\!H\!V\!\rangle \!\!+\!\!|\!V\!V\!H\!V\!V\!H\!\rangle),\nonumber\\
&{|\phi'\rangle}_0\!=\!\frac{1}{2}(|\!H\!H\!V\!H\!H\!V\!\rangle \!\!+\!\!|\!H\!V\!H\!H\!V\!H\!\rangle \!\!+\!\!|\!V\!H\!H\!V\!H\!H\!\rangle \!\!+\!\!|\!V\!V\!V\!V\!V\!V\!\rangle),\nonumber\\
&{|\phi\rangle}_1=\!\frac{1}{2}(|\!H\!H\!H\!H\!V\!V\!\rangle\!\!+\!\!|\!H\!V\!V\!H\!H\!H\!\rangle\!\!+\!\!|\!V\!H\!V\!V\!V\!H\!\rangle \!\!+\!\!|\!V\!V\!H\!V\!H\!V\!\rangle),\nonumber\\
&{|\phi'\rangle}_1\!=\!\frac{1}{2}(|\!H\!H\!V\!H\!V\!H\!\rangle\!\!+\!\!|\!H\!V\!H\!H\!H\!V\!\rangle\!\!+\!\!|\!V\!H\!H\!V\!V\!V\!\rangle \!\!+\!\!|\!V\!V\!V\!V\!H\!H\!\rangle),\nonumber\\
&{|\phi\rangle}_2=\!\frac{1}{2}(|\!H\!H\!H\!V\!H\!V\!\rangle \!\!+\!\!|\!H\!V\!V\!V\!V\!H\!\rangle \!\!+\!\!|\!V\!H\!V\!H\!H\!H\!\rangle \!\!+\!\!|\!V\!V\!H\!H\!V\!V\!\rangle),\nonumber\\
&{|\phi'\rangle}_2\!=\!\frac{1}{2}(|\!H\!H\!V\!V\!H\!H\!\rangle\!\!+\!\!|\!H\!V\!H\!V\!V\!V\!\rangle\!\!+\!\!|\!V\!H\!H\!H\!H\!V\!\rangle \!\!+\!\!|\!V\!V\!V\!H\!V\!H\!\rangle),\nonumber\\
&{|\phi\rangle}_3=\!\frac{1}{2}(|\!H\!H\!H\!V\!V\!H\!\rangle \!\!+\!\!|\!H\!V\!V\!V\!H\!V\!\rangle \!\!+\!\!|\!V\!H\!V\!H\!V\!V\!\rangle \!\!+\!\!|\!V\!V\!H\!H\!H\!H\!\rangle ),\nonumber\\
&{|\phi'\rangle}_3\!=\!\frac{1}{2}(|\!H\!H\!V\!V\!V\!V\!\rangle \!\!+\!\!|\!H\!V\!H\!V\!H\!H\!\rangle \!\!+\!\!|\!V\!H\!H\!H\!V\!H\!\rangle \!\!+\!\!|\!V\!V\!V\!H\!H\!V\!\rangle).
\end{align}
It is similar to the case where the initial entangled ensembles are the same \cite{36-pra-deng-mepp}, namely, ${^1\!}P_0\!=\!{^2\!}P_0$. ${|\phi\rangle }_i$ and ${|\phi'\rangle}_i$ ($i=1,2,3$) can be transformed into ${|\phi\rangle}_0$ and ${|\phi'\rangle}_0$ respectively with the local unitary operations in Table \ref{table2a}, here $I$ and $\sigma_x$ represent the identity operation and the bit-flip operation respectively. 

The successful purification is to pick up the even number of odd parity, namely, eee, eoo, oeo, or ooe, which correspond to the identity combinations. In this way, the parties can obtain a six-photon entangled system which is in the states ${|\phi\rangle }_0$ and ${|\phi'\rangle} _0$ with the probabilities $\frac{{^1\!}P_0{^2\!}P_0}{{^1\!}P_0{^2\!}P_0\!+\!{^1\!}P_1{^2\!}P_1}$ and $\frac{{^1\!}P_1{^2\!}P_1}{{^1\!}P_0{^2\!}P_0\!+\!{^1\!}P_1{^2\!}P_1}$ respectively. For the six-photon entangled systems, they can measure the three photons $A_2B_2C_2$ with the basis $X\!=\!\left \{|\pm\rangle \right \}$, 
and perform a phase-flip operation $\sigma_z$ on the specific qubits of system 1 if the corresponding qubits of system 2 is $|-\rangle $.
For example, if the result is $|--+\rangle $, then they need to perform $\sigma_z$ on $A_1$ and $B_1$. 
For all $2^3$ kinds of combinations ${|\pm\rangle }\otimes{|\pm\rangle }\otimes{|\pm\rangle } $,
the parties will eventually obtain the same new mixed states, which is composed of the states $|\Psi^+\rangle $ and $|\Psi^-\rangle $ with the probabilities $\frac{{^1\!}P_0{^2\!}P_0}{{^1\!}P_0{^2\!}P_0\!+\!{^1\!}P_1{^2\!}P_1}$ and $\frac{{^1\!}P_1{^2\!}P_1}{{^1\!}P_0{^2\!}P_0\!+\!{^1\!}P_1{^2\!}P_1}$ respectively.
Finally, with a Hadamard operation on each photon, the new systems obtained by the parties are
\begin{align}\label{eq55}
\rho'=P'_0|\Phi^+\rangle \langle \Phi^+|+P'_1|\Phi^-\rangle \langle \Phi^-|,
\end{align}
where
\begin{align}\label{eq56}
P'_0=\frac{{^1\!}P_0{^2\!}P_0}{{^1\!}P_0{^2\!}P_0+{^1\!}P_1{^2\!}P_1}.
\end{align}
Obviously, in order to satisfy $P'_0\!>\!{^1}P_0$, ${^2}P_0\!>\!1/2$ is needed. The lower bound of ${^2\!}P_0$ is $1/2$, which is completely consistent with the condition when the initial ensembles are the same \cite{36-pra-deng-mepp}. In addition, one can find that the fidelity $P'_0$ of the new systems $\rho'$ in Eq. \eqref{eq56} is exactly the same as that in the case of two different initial ensembles composed of two-photon systems with only bit-flip or phase-flip errors \cite{13-oe-diffepp}.

Interestingly, different from the case of the same initial ensembles, we show that there is also residual entanglement in the cross combinations if ${^1\!}P_0\!\ne\!{^2\!}P_0$,
which means the discarded item can also be reused. For the cross combinations $|\Psi^+\rangle \otimes|\Psi^-\rangle $ and $|\Psi^-\rangle \otimes|\Psi^+\rangle $, the relationship among the outcomes of QNDs measured by three parties Alice, Bob and Charlie, the states of the system composed of the six photons $A_1B_1C_1A_2B_2C_2$ and the probabilities are shown in Table \ref{table2b}, where
\begin{align}\label{eq57}
&{|\psi\rangle}_0=\!\frac{1}{2}(|\!H\!H\!H\!V\!V\!V\!\rangle \!\!+\!\!|\!H\!V\!V\!V\!H\!H\!\rangle \!\!+\!\!|\!V\!H\!V\!H\!V\!H\!\rangle \!\!+\!\!|\!V\!V\!H\!H\!H\!V\!\rangle ),\nonumber\\
&|\psi'\rangle _0\!=\!\frac{1}{2}(|\!H\!H\!V\!V\!V\!H\!\rangle \!\!+\!\!|\!H\!V\!H\!V\!H\!V\!\rangle \!\!+\!\!|\!V\!H\!H\!H\!V\!V\!\rangle \!\!+\!\!|\!V\!V\!V\!H\!H\!H\!\rangle ),\nonumber\\
&{|\psi\rangle} _1=\!\frac{1}{2}(|\!H\!H\!H\!V\!H\!H\!\rangle \!\!+\!\!|\!H\!V\!V\!V\!V\!V\!\rangle \!\!+\!\!|\!V\!H\!V\!H\!H\!V\!\rangle \!\!+\!\!|\!V\!V\!H\!H\!V\!H\!\rangle ),\nonumber\\
&|\psi'\rangle _1\!=\!\frac{1}{2}(|\!H\!H\!V\!V\!H\!V\!\rangle \!\!+\!\!|\!H\!V\!H\!V\!V\!H\!\rangle \!\!+\!\!|\!V\!H\!H\!H\!H\!H\!\rangle \!\!+\!\!|\!V\!V\!V\!H\!V\!V\!\rangle ),\nonumber\\
&{|\psi\rangle} _2=\!\frac{1}{2}(|\!H\!H\!H\!H\!V\!H\!\rangle \!\!+\!\!|\!H\!V\!V\!H\!H\!V\!\rangle \!\!+\!\!|\!V\!H\!V\!V\!V\!V\!\rangle \!\!+\!\!|\!V\!V\!H\!V\!H\!H\!\rangle ),\nonumber\\
&|\psi'\rangle _2\!=\!\frac{1}{2}(|\!H\!H\!V\!H\!V\!V\!\rangle \!\!+\!\!|\!H\!V\!H\!H\!H\!H\!\rangle \!\!+\!\!|\!V\!H\!H\!V\!V\!H\!\rangle \!\!+\!\!|\!V\!V\!V\!V\!H\!V\!\rangle ),\nonumber\\
&{|\psi\rangle} _3=\!\frac{1}{2}(|\!H\!H\!H\!H\!H\!V\!\rangle \!\!+\!\!|\!H\!V\!V\!H\!V\!H\!\rangle \!\!+\!\!|\!V\!H\!V\!V\!H\!H\!\rangle \!\!+\!\!|\!V\!V\!H\!V\!V\!V\!\rangle ),\nonumber\\
&|\psi'\rangle _3\!=\!\frac{1}{2}(|\!H\!H\!V\!H\!H\!H\!\rangle \!\!+\!\!|\!H\!V\!H\!H\!V\!V\!\rangle \!\!+\!\!|\!V\!H\!H\!V\!H\!V\!\rangle \!\!+\!\!|\!V\!V\!V\!V\!V\!H\!\rangle ).
\end{align}
That is, the instances that the number of odd numbers after parity check is odd, namely, ooo, oee, eoe or eeo, correspond to the cross combinations. With some local unitary operations, they can obtain a six-photon entangled system in the states $|\psi\rangle _0$ and $|\psi'\rangle _0$ with the probabilities $\frac{{^1\!}P_0{^2\!}P_1}{{^1\!}P_0{^2\!}P_1\!+\!{^1\!}P_1{^2\!}P_0}$ and $\frac{{^1\!}P_1{^2\!}P_0}{{^1\!}P_0{^2\!}P_1\!+\!{^1\!}P_1{^2\!}P_0}$ respectively. After the same operations as above, they can finally get the new mixed entangled state
\begin{align}\label{eq58}
\rho''=P''_0|\Phi^+\rangle \langle \Phi^+|+P''_1|\Phi^-\rangle \langle \Phi^-|,
\end{align}
where
\begin{align}\label{eq59}
P''_0=\frac{{^1\!}P_0{^2\!}P_1}{{^1\!}P_0{^2\!}P_1+{^1\!}P_1{^2\!}P_0}.
\end{align}
Here, the parties successfully obtain the residual entanglement from the cross combinations, which is not available in the case of the same initial ensembles \cite{36-pra-deng-mepp} (${^1\!}P_0\!=\!{^2\!}P_0$). 
In addition, The fidelity of the ensemble $\rho''$ usually discarded satisfies $P''_0>1/2$ if ${^1\!}P_0>{^2\!}P_0(>1/2)$, that is, if there is only phase-flip errors in three-photon systems, the residual entanglement always exists if the fidelity of two initial mixed state are different.   

To obtain a higher fidelity, they can perform the purification in a second round with two copies of mixed states $\rho''$, and if only considering the identity combinations, then the fidelity of the new mixed state $P'''_0$ is 
\begin{align}\label{eq60}
P'''_0&=\frac{{P''_0}^2}{{P''_0}^2+{(1-P''_0)}^2}\nonumber\\
&=\frac{{{^1\!}P_0}^2{(1-{^2\!}P_0)}^2}{{{^1\!}P_0}^2{(1-{^2\!}P_0)}^2+{{^2\!}P_0}^2{(1-{^1\!}P_0)}^2}.
\end{align}
In the second purification, in order to satisfy $P'''_0\!>\!{^1\!}P_0$, namely, to obtain a higher fidelity mixed state with the residual entanglement, we require
\begin{align}\label{eq61}
\frac{{^1\!}P_0}{1-{^1\!}P_0}> \frac{{{^2\!}P_0}^2}{{(1-{^2\!}P_0)}^2}.
\end{align}
This is completely consistent with the conclusion of Ref. \cite{13-oe-diffepp}, which is the same as the case of the different two-photon ensembles with only bit-flip or phase-flip errors.

It is straightforward to extend the above three-photon case to the N-photon GHZ state. For a phase flip error, we also perform a Hadamard operation on all photons to transform the states ${|\Phi_0^+\rangle} _N$ and ${|\Phi_0^-\rangle} _N$ in Eq. \eqref{eq42} to ${|\Psi^+\rangle} _N$ and ${|\Psi^-\rangle }_N$, respectively. Here 
\begin{align}\label{eq62}
{|\Psi^+\rangle}_N=&\frac{1}{2^{\frac{N+1}{2}}}[(|H\rangle +|V\rangle )(|H\rangle +|V\rangle )...(|H\rangle +|V\rangle )\nonumber\\
&+(|H\rangle -|V\rangle )(|H\rangle -|V\rangle )...(|H\rangle -|V\rangle )]\nonumber\\
=&\frac{1}{2^{\frac{N-1}{2}}}\sum_{\substack{i,j,...,k\in\left\{0,1\right\},\\2|(i+j+...+k)}}|ij...k\rangle ,\nonumber\\
{|\Psi^-\rangle} _N=&\frac{1}{2^{\frac{N+1}{2}}}[(|H\rangle +|V\rangle )(|H\rangle +|V\rangle )...(|H\rangle +|V\rangle )\nonumber\\
&-(|H\rangle -|V\rangle )(|H\rangle -|V\rangle )...(|H\rangle -|V\rangle )]\nonumber\\
=&\frac{1}{2^{\frac{N-1}{2}}}\sum_{\substack{i,j,...,k\in\left\{0,1\right\},\\2\nmid(i+j+...+k)}}|ij...k\rangle.
\end{align}
Here  $2\!\mid\!(i+j+...+k)$ and $2\!\nmid\!(i+j+...+k)$ respectively represent that the number of $|V\rangle $ in each item is even or odd. 
Due to the lower symmetry in N-photon GHZ states, we cannot utilize Hadamard operations to realize the conversion between a phase-flip error and a single bit-flip error, so as to directly copy the process of correction for bit-flip errors. But likewise, the parity characteristics can be used to achieve purification. 
We can also obtain a mixed state, written as
\begin{align}\label{eq63}
{^1\!}{\rho}_N={^1\!}P_0{|\Psi^+\rangle} _N\langle \Psi^+|+{^1\!}P_1{|\Psi^-\rangle }_N\langle \Psi^-|,\nonumber\\
{^2\!}{\rho}_N={^2\!}P_0{|\Psi^+\rangle }_N\langle \Psi^+|+{^2\!}P_1{|\Psi^-\rangle }_N\langle \Psi^-|,
\end{align}
after performing the Hadamard operations.
After each party performs the parity-check measurement on the two photons respectively, the instances are divided into two groups according to the parity of odd numbers in parity-check result. 
If the parity is even, it corresponds to the identity combinations ${|\Psi^+\rangle} _N\otimes{|\Psi^+\rangle }_N$ and ${|\Psi^-\rangle }_N\otimes{|\Psi^-\rangle} _N$. For the other case, it corresponds to the cross combinations ${|\Psi^+\rangle} _N\otimes{|\Psi^-\rangle }_N$ and ${|\Psi^-\rangle }_N\otimes{|\Psi^+\rangle} _N$, which is usually discarded. After some unitary operations in both cases, the 2N-photon system can be in the states
\begin{align}\label{eq64}
|\phi\rangle &=\frac{1}{2^{\frac{N-1}{2}}}\sum_{\substack{i,j,...,k\in\left\{0,1\right\},\\2|(i+j+...+k)}}|ij...k\rangle _1\otimes|ij...k\rangle _2,\nonumber\\
|\phi'\rangle &=\frac{1}{2^{\frac{N-1}{2}}}\sum_{\substack{i,j,...,k\in\left\{0,1\right\},\\2\nmid(i+j+...+k)}}|ij...k\rangle _1\otimes|ij...k\rangle _2,
\end{align}
with the probabilities $\frac{{^1\!}P_0{^2\!}P_0}{{^1\!}P_0{^2\!}P_0\!+\!{^1\!}P_1{^2\!}P_1}$ and $\frac{{^1\!}P_1{^2\!}P_1}{{^1\!}P_0{^2\!}P_0\!+\!{^1\!}P_1{^2\!}P_1}$, or in the states
\begin{align}\label{eq641}
|\psi\rangle &=\frac{1}{2^{\frac{N-1}{2}}}\sum_{\substack{i,j,...,k\in\left\{0,1\right\},\\2|(i+j+...+k)}}|ij...k\rangle _1\otimes|\bar{i}\bar{j}...\bar{k}\rangle _2,\nonumber\\
|\psi'\rangle &=\frac{1}{2^{\frac{N-1}{2}}}\sum_{\substack{i,j,...,k\in\left\{0,1\right\},\\2\nmid(i+j+...+k)}}|ij...k\rangle _1\otimes|\bar{i}\bar{j}...\bar{k}\rangle _2,
\end{align}
with the probabilities $\frac{{^1\!}P_0{^2\!}P_1}{{^1\!}P_0{^2\!}P_1\!+\!{^1\!}P_1{^2\!}P_0}$ and $\frac{{^1\!}P_1{^2\!}P_0}{{^1\!}P_0{^2\!}P_1\!+\!{^1\!}P_1{^2\!}P_0}$, respectively.
Then, the parties measure the photons of system 2 with the basis $X\!=\!\left \{|\pm\rangle \right \}$, and will obtain a new system in the mixed state which is composed of the states ${|\Psi^+\rangle} _N$ and ${|\Psi^-\rangle} _N$ with the probabilities $\frac{{^1\!}P_0{^2\!}P_0}{{^1\!}P_0{^2\!}P_0\!+\!{^1\!}P_1{^2\!}P_1}$ and $\frac{{^1\!}P_1{^2\!}P_1}{{^1\!}P_0{^2\!}P_0\!+\!{^1\!}P_1{^2\!}P_1}$ respectively. 
Significantly, if some qubits of system 2 is measured to be $|-\rangle $, then the parties should perform a phase-flip operation $\sigma_z$ on the corresponding qubits of system 1. For instance, they should perform $\sigma_z$ on the 1st,..., and Nth qubits of system 1 respectively if the measurement result is $|-+...+-\rangle $. 

Finally, after a Hadamard operation, the parties will finally obtain the new mixed state with the same fidelity $P'_0$ and $P''_0$ in Eq. \eqref{eq56} and  Eq. \eqref{eq59}.
Therefore, if only phase-flip error occurs in N-photon systems, the formulas of the new fidelities and the conditions for improving the fidelity are completely consistent, independent of N (N $\!\ge\!2$). And due to the special high symmetry of Bell states, the two-photon case is a special example, in which a phase-flip error and a single bit-flip error can be equivalently converted into each other through the Hadamard operation.
In addition, there is always residual entanglement that can be reused in the discarded cross combinations when the ensembles are different.

\section{Discussion}\label{GOMEPP}
To sum up, the MEPP for different initial fidelities has been described. We first discussed the MEPP for bit-flip errors using the approach of entanglement link, i. e., P1. We also showed that the cross combinations still have entanglement which can be reused directly, i. e., the approach of residual entanglement purification, named P1$^\prime$. Finally, we also discussed the MEPP for phase-flip errors. Interestingly, we still can exploit the approach of residual entanglement purification for cross combinations. Therefore, it is worth to discuss the optimal entanglement purification for multipartite entanglement. 

Before that, we first make a comparative analysis between approach P1$^\prime$ and P1, and here we consider the case of symmetric three-photon systems  for simplicity. The fidelities of the new ensembles mentioned above can be summarized as follows, for the case of bit-flip error, there are
\begin{align}\label{eq69-71}
F_0'=&\frac{3{^1\!}F_0{^2\!}F_0}{1-{^1\!}F_0-{^2\!}F_0+4{^1\!}F_0{^2\!}F_0}\nonumber\\
&\mathop{\longrightarrow }\limits^{{^1\!}F_0={^2\!}F_0=F_0}\frac{3{F_0}^2}{4{F_0}^2-2F_0+1},\\
F_0^{\,t}=&\frac{9{({^1\!}F_0+{^2\!}F_0-2{^1\!}F_0{^2\!}F_0)}^2}{{({^1\!}F_0+{^2\!}F_0-4{^1\!}F_0{^2\!}F_0+2)}^2}\nonumber\\
&\mathop{\longrightarrow }\limits^{{^1\!}F_0={^2\!}F_0=F_0}\frac{9{F_0}^2}{4{F_0}^2+4F_0+1},\\
F''_0=&\frac{3{^1\!}F_0(1-{^2\!}F_0)}{{^1\!}F_0+{^2\!}F_0-4{^1\!}F_0{^2\!}F_0+2}\nonumber\\
&\mathop{\longrightarrow }\limits^{{^1\!}F_0={^2\!}F_0=F_0}\frac{3F_0}{4F_0+2},
\end{align}
and
\begin{align}\label{eq72}
F'''_0=&\frac{9{{^1\!}F_0}^2{(1\!-\!{^2\!}F_0)}^2}{9{{^1\!}F_0}^2{(1\!-\!{^2\!}F_0)}^2\!+\!9{{^2\!}F_0}^2{(1\!-\!{^1\!}F_0)}^2\!+\!2{(1\!-\!{^1\!}F_0)}^2{(1\!-\!{^2\!}F_0)}^2}\nonumber\\
&\mathop{\longrightarrow }\limits^{{^1\!}F_0={^2\!}F_0=F_0}\frac{9{F_0}^2}{20{F_0}^2-4F_0+2}.
\end{align}
For phase-flip error correction, there are
\begin{align}\label{eq73-74}
P'_0=&\frac{{^1\!}P_0{^2\!}P_0}{{^1\!}P_0{^2\!}P_0+{^1\!}P_1{^2\!}P_1}\nonumber\\
&\mathop{\longrightarrow }\limits^{{^1\!}P_0={^2\!}P_0=P_0}\frac{{P_0}^2}{2{P_0}^2-2P_0+1},\\
P'''_0=&\frac{{{^1\!}P_0}^2{(1-{^2\!}P_0)}^2}{{{^1\!}P_0}^2{(1-{^2\!}P_0)}^2+{{^2\!}P_0}^2{(1-{^1\!}P_0)}^2}.
\end{align}
Meanwhile, in the cases of only bit-flip errors or phase-flip errors, we have previously given the conditions that the parties obtain the higher-fidelity new systems, namely, $F'_0\!>\!{^1\!}F_0$, $F^{t}_0\!>\!{^1\!}F_0$, $F'''_0\!>\!{^1\!}F_0$, $P_0'\!>\!{^1\!}P_0$ and $P'''_0\!>\!{^1\!}P_0$. Here we simplify and list them respectively in the case of symmetric initial three-photon entangled ensembles as
\begin{align}\label{eq75-79}
F_0'&> {^1\!}F_0\Longleftrightarrow {^2\!}F_0> \frac{1}{4},\\
F_0^{t}&> {^1\!}F_0\Longleftrightarrow \frac{9{({^1\!}F_0\!+\!{^2\!}F_0\!-\!2{^1\!}F_0{^2\!}F_0)}^2}{{({^1\!}F_0\!+\!{^2\!}F_0\!-\!4{^1\!}F_0{^2\!}F_0\!+\!2)}^2}> {^1\!}F_0,\\
F'''_0&> {^1\!}F_0\Longleftrightarrow \frac{9{^1\!}F_0}{1\!-\!{^1\!}F_0}> \frac{11{{^2\!}F_0}^2\!-\!4{^2\!}F_0\!+\!2}{{(1\!-\!{^2\!}F_0)}^2},\\
P_0'&> {^1\!}P_0\Longleftrightarrow {^2\!}P_0> \frac{1}{2},\\
P'''_0&> {^1\!}P_0\Longleftrightarrow \frac{{^1\!}P_0}{1-{^1\!}P_0}> \frac{{{^2\!}P_0}^2}{{(1-{^2\!}P_0)}^2}.
\end{align}

\begin{figure}[thp]
\centering
\includegraphics[width=0.4\textwidth]{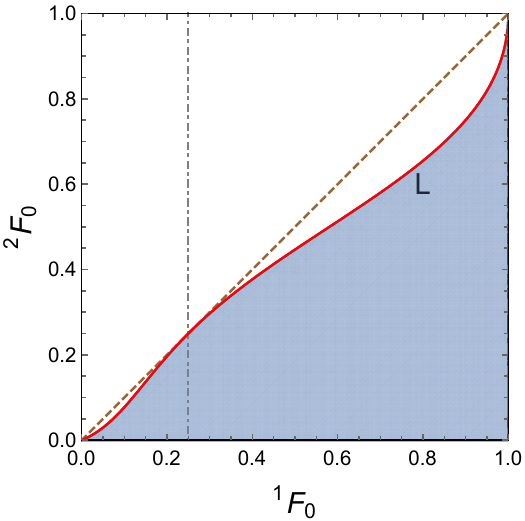}
\caption{The feasible region where $F'''_0\!>\!F^{t}_0$.}
\label{pic8}
\end{figure}

Significantly, in the process of correction for bit-flip errors, scheme P1 and scheme P1$^\prime$ consume the same entanglement resources. Specifically, these two schemes both require two pairs of entangled systems, where a pair of entangled systems are composed of one system of ensemble 1 and one system of ensemble 2, for obtaining the new fidelities $F^{t}_0$ and $F'''_0$ from the cross combinations, respectively.
In order to decide which scheme to choose, we need to compare the fidelity $F^{t}_0$ in scheme P1 and the fidelity $F'''_0$ in scheme P1$^\prime$. As shown in Fig. \ref{pic8}, we draw the region of $F'''_0\!>\!F^{t}_0$. One can find that when ${^1}F_0\!=\!{^2}F_0\!=\!F_0$, $F'''_0\!\le \!F^{t}_0$ always holds (the equality holds iff $F_0\!=\!1/4$, where the initial ensembles are in a mixed state with equal probability and cannot be purified, wherefore any part of MEPP cannot work), and the parties should only choose scheme P1 in this special case. Interestingly, if the initial two mixed states are different, that is, ${^1}F_0\!\ne\!{^2}F_0$, it's more likely that $F'''_0$ is greater than $F^{t}_0$. That is to say, the fidelity obtained by choosing scheme P1$^\prime$ is optimal. For instance, if ${^1}F_0\!=\!0.8$, ${^2}F_0\!=\!0.6$, there is $F'''_0\!=\!0.866\!>\!F^{t}_0\!=\!0.795$. 
One can find that scheme P1$^\prime$ succeeds while scheme P1 fails. The red critical curve L indicates $F'''_0\!=\!F^{t}_0$.

\begin{figure}[thp]
\centering
\includegraphics[width=0.4\textwidth]{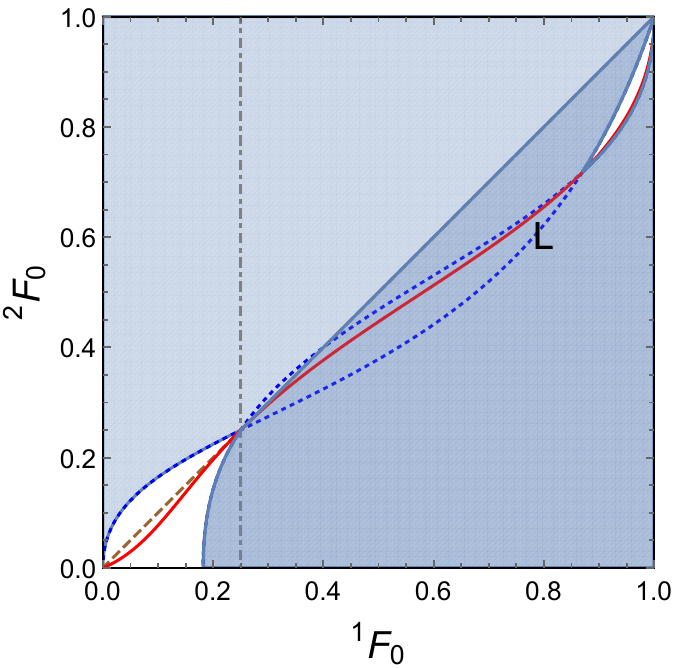}
\caption{The feasible region where the new fidelity satisfies $\max\{F^{t}_0,F'''_0\}\!>\!{^1\!}F_0$ in new scheme P1/P1$^\prime$. The red line L here is the same as Fig. \ref{pic8}, and the blue dotted lines are the boundary lines in Fig. \ref{pic3} and Fig. \ref{pic7}.}
\label{pic9}
\end{figure}

Therefore, in order to obtain the higher fidelity, we should combine scheme P1 and scheme P1$^\prime$ in the general case of different initial ensembles, and decide which scheme to use according to some known parameters in advance. In this way, with our new scheme P1/P1$^\prime$, the feasible region where the fidelity of new systems obtained from the cross combinations satisfies $\max\{F^{t}_0,F'''_0\}\!>\!{^1\!}F_0$ is shown in Fig. \ref{pic9}. We find that in most cases the parties can use P1/P1$^\prime$ to obtain the new states with fidelity greater than ${^1\!}F_0$ from the cross combinations.

\begin{figure}[thp]
\centering
\includegraphics[width=0.45\textwidth]{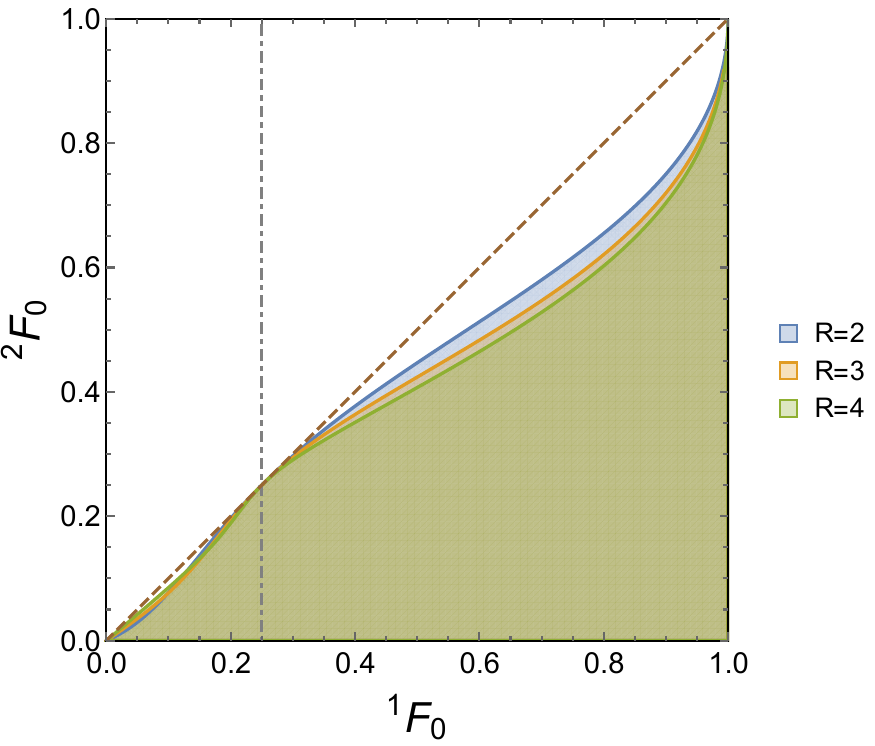}
\caption{The feasible region where $F'''^{(R)}_0\!>\!F^{t(R-2)}_0$, $R\!=\!2$, $3$, $4$, in the case of multiple rounds of purification.}
\label{pic10}
\end{figure}

Moreover, in consideration of multiple rounds of purification, we can also give the comparison between scheme P1 and scheme P1$^\prime$, as shown in Fig. \ref{pic10}. For example, in terms of $R$ rounds purification in scheme P1$^\prime$, the parties require $2^{R-1}$ pairs of entangled systems. Likewise, scheme P1, in which the two-photon systems are purified for $R\!-\!2$ rounds in advance, and then regenerate the three-photon systems with entanglement links, also needs to consume the same entanglement resources. Thus, the more general criterion is $F'''^{(R)}_0\!>\!F^{t(R-2)}_0$, here $R\!\ge\!2$. The numbers in brackets represent the number of rounds of purification in scheme P1$^\prime$ or that of two-photon purification in advance in scheme P1. If $R\!=\!2$, it's the case of two rounds of purification above, namely, $F'''_0\!>\!F^{t}_0$.

Lastly, we briefly provide the comparison of these two schemes in practical applications. 
In the case of N-photon, the scheme P1 requires sophisticated entanglement links, some of which even consume more entanglement resources compared with P1$^\prime$. While the scheme P1$^\prime$ only needs to expand the number of QNDs in Fig. \ref{pic1} instead of complex entanglement links and is easy to be extended to the N-photon case, which seems more feasible in practical application. Thus, the combination of P1 and P1$^\prime$ in practice requires consideration of numerous above factors.

In a more general case where bit-flip errors and phase-flip errors may occur in the meantime, our general optimized recurrence MEPP will contain two modules. One is P1/P1$^\prime$ corresponding to correction for bit-flip errors, and the other is P2 corresponding to correction for phase-flip errors. It's similar to that mentioned in Ref. \cite{earlymepp2,earlymepp3,36-pra-deng-mepp}, we don't have to execute the two modules in the fixed order of P1/P1$^\prime$,P2,P1/P1$^\prime$,P2... the choice between P1 or P1$^\prime$, and the order of P1/P1$^\prime$ and P2 can be arbitrary. 
Also, the feasibility, the efficiency, the multiple rounds of purification and even the allocation of entanglement resources need to be taken into account. Therefore, for multipartite systems with many parameters, we can't get a deterministic expression to describe the iteration of the fidelity of ensembles obtained. 
In practice, according to the known parameters of density matrices and the final expectation of fidelity, we can establish a multi-objective programming (a classical selector) with the goal of minimum execution steps, the maximum number of entanglement resources obtained (efficiency) and so on, through some optimization numerical methods \cite{optimization-epp1,optimization-epp2}, to give the optimal order of MEPP. If other new schemes are proposed in the future, they can also be included in the classical selector.

It is noteworthy that in our general optimized MEPP, the cross-Kerr nonlinearity is used to construct QNDs which has the functions of both parity-check gates and PNDs. Thus, with PNDs, this MEPP can work efficiently with photon-loss channels as well, different from those based on CNOT gates or linear optical elements. Also, in addition to the cross-Kerr nonlinearity, there are other methods to construct QNDs, such as quantum dots in optical cavities. 
Moreover, similar to Ref. \cite{36-pra-deng-mepp}, both schemes P1, P1$^\prime$ and P2 can work by replacing QNDs with CNOT gates, so that this MEPP can also works by replacing parity-check detectors with CNOT gates or replacing the polarization degree of freedom of N-photon systems with others. 
Whereas, if the parties replace their QNDs with polarizing beam splitters, current schemes P1, P1$^\prime$ and P2 will not work, because the cross combinations cannot be exploited.

\section{Conclusion}\label{dc}
We proposed an efficient MEPP for N-photon GHZ states with different initial fidelities, and give the detailed processes and the related analyses.
Besides the cases of successful purification, this MEPP actually utilize the discarded items which are usually regarded as a failure.
Our protocol contains two parts for bit-flip error correction. The first one is the conventional MEPP with different initial fidelity, corresponding successful cases. 
The second one includes two efficient approaches that can utilize discarded items, recycling purification with entanglement link and direct residual entanglement purification, named P1 and P1$^\prime$, respectively. 
In the approach P1, by measuring some photons, the discarded items can become entangled N$^\prime$-photon (2 $\le$ N$^\prime < $ N) subsystems which can be reused through entanglement links to generate a N-photon entangled state. 
The other approach P1$^\prime$ is to directly utilize residual entanglement of the discarded items to improve the yield of high quality entangled states.
We also make a comparison between two approaches. In most cases the approach of direct residual purification is optimal, because it not only may obtain a higher fidelity entangled state but also it is more feasible in experiment.
In addition, after performing the purification for the phase-flip error correction, the discarded items still have available residual entanglement in the case of different input states. This whole approach is called P2.
We discuss the principle of our MEPP in detail for three-photon systems, and the conclusions can be generalized to the case of N-photon systems.
Considering numerous factors in practice, we can establish a multi-objective programming according to the goals and the known initial state information, to give the optimal order of MEPP through some optimization numerical methods. 
Based on the above, our general optimized MEPP has greater advantages than all previous MEPPs and it may have potential applications in the long-distance quantum communications and quantum networks.

\section*{Acknowledgement}
This work is supported by the National Natural Science Foundation of China under Grant  Nos. 11974189, 12175106 and 92365110.

\section*{Appendix : Efficiencies and average fidelities of the entanglement purification schemes P1 and P1$^\prime$}\label{eff}
\setcounter{equation}{0}
\setcounter{subsection}{0}
\renewcommand{\theequation}{A\arabic{equation}}
The schemes P1 and P1$^\prime$ has been introduced in Sec. \ref{bit}, and here we will make a simple analysis of their efficiencies and average fidelities. Similar to Ref. \cite{36-pra-deng-mepp}, the efficiency of an MEPP $Y$ is defined as the probability (i.e., the yield) that the parties can obtain a high fidelity entangled multi-photon system from a pair of multi-photon systems transmitted over a noisy channel without loss, which reflects the resource utilization rate of MEPP. The efficiencies of MEPP $Y_{P1}$ and $Y_{P1^\prime}$ depend on the parameters ${^\alpha\!}F_i$. For simplicity, we  discuss the case of symmetric initial ensembles below.

Firstly, for the scheme P1, we denote the efficiency of the purification scheme of the identity combinations ${|\Phi_i^+\rangle }_1\otimes{|\Phi_i^+\rangle }_2$ as $Y_i$, obviously it's also the probability that the pair of three-photon systems are in the identity combinations. In the above situation, the efficiency of the three-photon MEPP $Y_i$ is
\begin{align}\label{eq29}
Y_i&=P_{identity}\nonumber\\
&={^1\!}F_0{^2\!}F_0+{^1\!}F_1{^2\!}F_1+{^1\!}F_2{^2\!\!}F_2+{^1\!}F_3{^2\!}F_3\nonumber\\
&=\frac{1-{^1\!}F_0-{^2\!}F_0+4{^1\!}F_0{^2\!}F_0}{3}.
\end{align}
The probability of cross combinations $P_ {cross}$, that is also the probability of the parties obtaining two-photon pairs from a pair of three-photon systems $P_{3\to2}$, which can be written as
\begin{align}\label{eq30}
P_{3\to2}&=P_{cross}\nonumber\\
&=\sum_{i\ne j=0}^{3} {^1\!}F_i{^2\!}F_j\nonumber\\
&=\frac{2+{^1\!}F_0+{^2\!}F_0-4{^1\!}F_0{^2\!}F_0}{3}.
\end{align}
Here $P_{cross}\!+\!P_{identity}\!=\!1$. The efficiency of extracting three-photon entangled system from a pair of two-photon entangled systems with the entanglement link $Y_{2\to3}$ is only half of it, namely,
\begin{align}\label{eq31}
Y_{2\to3}&=\frac{1}{2}P_{3\to2}.
\end{align}
Then we can get the efficiency of scheme P1 with the entanglement link:
\begin{align}\label{eq32}
Y_{P1}=Y_i+Y_{2\to3}=\frac{2-\frac{1}{2}({^1}F_0+{^2}F_0)+2{^1}F_0{^2}F_0}{3}.
\end{align}

\begin{figure}[htp]
\includegraphics[width=.5\textwidth]{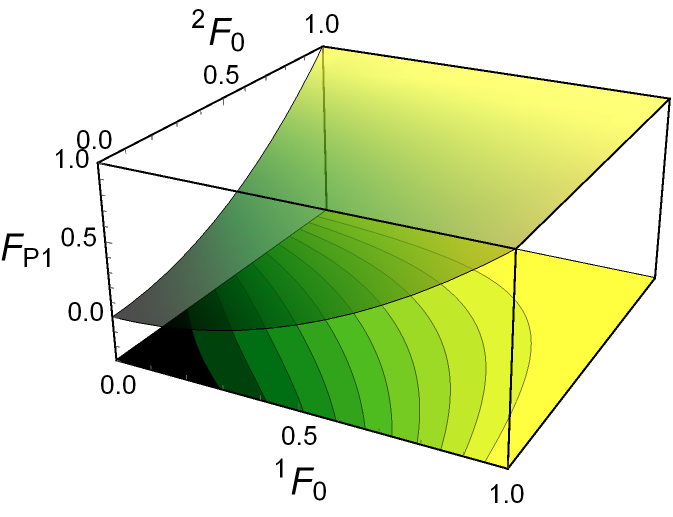}
\caption{The average fidelity $F_{P1}$ of scheme P1.}
\label{pic11}
\end{figure}

Also, in the case of symmetric initial entangled ensembles, we can obtain
\begin{align}\label{eq33}
F_i={F}_0'=\frac{3{^1\!}F_0{^2\!}F_0}{1-{^1\!}F_0-{^2\!}F_0+4{^1\!}F_0{^2\!}F_0}
\end{align}
and
\begin{align}\label{eq34}
F_{2\to3}=F_0^{\,t}=\frac{9{({^1\!}F_0+{^2\!}F_0-2{^1\!}F_0{^2\!}F_0)}^2}{{({^1\!}F_0+{^2\!}F_0-4{^1\!}F_0{^2\!}F_0+2)}^2},
\end{align}
according to Eq. \eqref{eq6} and \eqref{eq17}. Thus, the average fidelities of scheme P1 can be calculated as
\begin{align}\label{eq35}
F_{P1}&=\frac{F_iY_i+F_{2\to3}Y_{2\to3}}{Y_{P1}}\nonumber\\
&=\frac{6{^1\!}F_0{^2\!}F_0}{4-{^1\!}F_0-{^2\!}F_0+4{^1\!}F_0{^2\!}F_0}+\nonumber\\
&\frac{9{({^1\!}F_0\!+\!{^2\!}F_0\!-\!2{^1\!}F_0{^2\!}F_0)}^2(2\!+\!{^1\!}F_0\!+\!{^2\!}F_0\!-\!4{^1\!}F_0{^2\!}F_0)}{{({^1\!}F_0\!+\!{^2\!}F_0\!-\!4{^1\!}F_0{^2\!}F_0\!+\!2)}^2(4\!-\!{^1\!}F_0\!-\!{^2\!}F_0\!+\!4{^1\!}F_0{^2\!}F_0)}.
\end{align}
In Fig. \ref{pic11}, we drawn the $F_{P1}$ altered with ${^1\!}F_0$ and ${^2\!}F_0$ in Eq. \eqref{eq35}. It's easy to prove that, if ${^1\!}F_0\!=\!{^2\!}F_0\!=\!F_0$, Eq. \eqref{eq35} will revert to Eq. $(36)$ in Ref. \cite{36-pra-deng-mepp}, which is the special case of the same initial ensembles. 

After introducing the efficiency and the average fidelity of scheme P1, we will deduce them in the case of scheme P1$^\prime$. Before calculating the efficiency, we first make a simple analysis of the relationship between $F_0'$, $F_0''$ and ${^1\!}F_0$ in general three-photon systems. The efficiency of the identity combinations $Y_i$ is obviously the same as Eq. \eqref{eq29}, and the efficiency of cross combinations $Y_c$ equals $P_{cross}$, thus,
\begin{align}\label{eq36}
Y_i&=P_{identity}={^1\!}F_0{^2\!}F_0+{^1\!}F_1{^2\!}F_1+{^1\!}F_2{^2\!}F_2+{^1\!}F_3{^2\!}F_3,\nonumber\\
Y_c&=P_{cross}=\sum_{i\ne j=0}^{3} {^1\!}F_i{^2\!}F_j,
\end{align}
where $Y_i\!+\!Y_c\!=\!1$ always holds, and the probability (efficiency) of each parity-check result after QNDs meets
\begin{align}\label{eq37}
&Y_{eoe/oeo}\!=\!P_{eoe/oeo}\!=\!{^1\!}F_0{^2\!}F_2\!+\!{^1\!}F_2{^2\!}F_0\!+\!{^1\!}F_1{^2\!}F_3\!+\!{^1\!}F_3{^2\!}F_1,\nonumber\\
&Y_{oee/eoo}\!=\!P_{oee/eoo}\!=\!{^1\!}F_0{^2\!}F_1\!+\!{^1\!}F_1{^2\!}F_0\!+\!{^1\!}F_2{^2\!}F_3\!+\!{^1\!}F_3{^2\!}F_2,\nonumber\\
&Y_{eeo/ooe}\!=\!P_{eeo/ooe}\!=\!{^1\!}F_0{^2\!}F_3\!+\!{^1\!}F_3{^2\!}F_0\!+\!{^1\!}F_1{^2\!}F_2\!+\!{^1\!}F_2{^2\!}F_1,\nonumber\\
&Y_c=Y_{eoe/oeo}+Y_{eeo/ooe}+Y_{oee/eoo}.
\end{align}
Therefore, before the second purification of the new systems obtained from the cross combinations, the average fidelity of the four kinds of new systems ${\rho}_{(eoe/oeo)}''$, ${\rho}_{(eeo/ooe)}''$, ${\rho}_{(oee/eoo)}''$ and ${\rho}_{(eee/ooo)}'$ corresponding to the four parity-check results is
\begin{align}\label{eq38}
&\frac{F_iY_i\!+\!F_{0(eoe\!/\!oeo)}''\!Y_{eoe\!/\!oeo}\!+\!F_{0(eeo\!/\!ooe)}''\!Y_{eeo\!/\!ooe}\!+\!F_{0(oee\!/\!eoo)}''\!Y_{oee\!/\!eoo}}{Y_i\!+\!Y_{eoe/oeo}\!+\!Y_{eeo/ooe}\!+\!Y_{oee/eoo}}\nonumber\\&={^1\!}F_0({^2\!}F_0+{^2\!}F_1+{^2\!}F_2+{^2\!}F_3)={^1\!}F_0.
\end{align}
One can find that the average fidelity of all new systems obtained from different parity-check results is equal to ${^1\!}F_0$. Obviously when we get some high-fidelity new systems $\rho'$ from the identity combinations, there is $F^{'}_0\!>\! {^1\!}F_0$, then on average, the average fidelity $F''_0$ of the new systems $\rho''$ obtained from the cross combinations will be lower than ${^1\!}F_0$. If the average fidelity of all new systems $\rho''$ obtained from the cross combinations is higher than ${^1\!}F_0$, then the fidelity of the new systems obtained from the identity combinations will become $F'_0\!<\! {^1\!}F_0$. Actually, these two cases occur to take ${^2\!}F_0=1/4$ as the dividing point when the initial entangled ensembles are symmetric. For the special case of two-photon entanglement, the dividing point of the fidelity relationship of the new systems obtained from identity combinations and cross combinations is ${^2\!}F_0=1/2$ \cite{13-oe-diffepp}. We can easily prove that it's also true for the general case of N-photon systems, and the dividing point is ${^2\!}F_0=1/2^{N-1}$.

From another perspective, the treatment of the cross combinations is basically the same as that of the identity combinations, and the latter is a symmetrical purification scheme for the source ensemble ${^1\!}\rho$ and the target ensemble ${^2\!}\rho$ in the light of Eq. \eqref{eq6}, which means that the new system with high fidelity requires ${^1\!}F_0$ and ${^2\!}F_0$ close to 1. 
Contrasting Fig. \ref{pic4} with Fig. \ref{pic6}, 
according to the gradient direction of contours, it can be intuitively found that in scheme P1$^\prime$, the larger ${^1\!}F_0$ is compared with ${^2\!}F_0$, the higher the fidelity of the new systems obtained from the cross combinations is. 
In addition, we have understood the mathematical reason why scheme P1$^\prime$ requires a large difference between ${^2\!}F_0$ and ${^1\!}F_0$ from the analysis of Eq. \eqref{eq26} and \eqref{eq27}, whereas it actually contains the physical connotation of entanglement transformation. 
That is, the requirements of the identity combinations and the cross combinations for the initial fidelities of ensembles are essentially opposite, which is the result of entanglement transformation. In fact, we can see from the above that the parties cannot obtain the higher fidelity $F'_0$ from the identity combinations and the higher fidelity $F''_0$ from the cross combinations at the same time. Only one of the two can be improved higher. Usually we choose to obtain higher $F'_0$ in all MEPPs, so the cross combinations with low-fidelity can only be discarded in conventional MEPPs, or reused by the second round of purification or the entanglement links in our MEPP.
In essence, the EPP is actually the entanglement transformation process \cite{13-oe-diffepp}, 
so there is a contradiction between the requirements of the identity combinations and the cross combinations for the initial fidelities, which is the result of entanglement transformation.

\begin{figure}[t]
\centering
\includegraphics[width=0.5\textwidth]{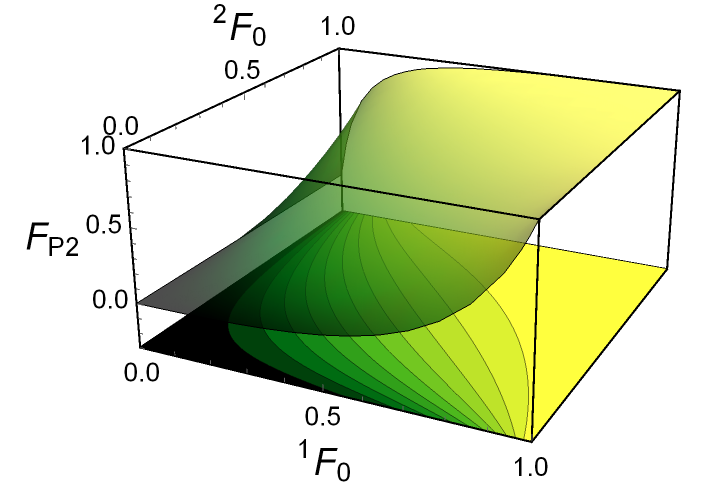}
\caption{The average fidelity $F_{P1'}$ of scheme P1$^\prime$.}
\label{pic12}
\end{figure}

Next, we analyse the efficiency of P1$^\prime$. In scheme P1$^\prime$, each system $\rho''$ obtains the maximum fidelity with the operation $\sigma_x$ and then perform the second round of purification. Here we only consider the identity combinations in the second round of purification, while the residual entanglement of the cross combinations are not included and will be calculated in the next round of purification. The efficiencies that the parties obtain $\rho''_{eoe/oeo}$, $\rho''_{oee/eoo}$ and $\rho''_{eeo/ooe}$ from a pair of three-photon systems are $Y_{eoe/oeo}$, $Y_{oee/eoo}$ and $Y_{eeo/ooe}$, respectively. Meanwhile, the probabilities of them evolving into $\rho'''_{eoe/oeo}$, $\rho'''_{oee/eoo}$ and $\rho'''_{eeo/ooe}$ after the second round of purification are respectively $P'_{i(eoe/oeo)}$, $P'_{i(oee/eoo)}$ and $P'_{i(eeo/ooe)}$, namely,
\begin{align}\label{eq39}
P'_{i(eoe/oeo)}&\!=\!{({^1\!}F_0{^2\!}F_2)}^2\!+\!{({^1\!}F_2{^2\!}F_0)}^2\!+\!{({^1\!}F_1{^2\!}F_3)}^2\!+\!{({^1\!}F_3{^2\!}F_1)}^2,\nonumber\\
P'_{i(oee/eoo)}&\!=\!{({^1\!}F_0{^2\!}F_1)}^2\!+\!{({^1\!}F_1{^2\!}F_0)}^2\!+\!{({^1\!}F_2{^2\!}F_3)}^2\!+\!{({^1\!}F_3{^2\!}F_2)}^2,\nonumber\\
P'_{i(eeo/ooe)}&\!=\!{({^1\!}F_0{^2\!}F_3)}^2\!+\!{({^1\!}F_3{^2\!}F_0)}^2\!+\!{({^1\!}F_1{^2\!}F_2)}^2\!+\!{({^1\!}F_2{^2\!}F_1)}^2.
\end{align}
In the second round of purification, we need to extract the high-fidelity three-photon entangled system from the six-photon entangled system, so the efficiency is only half of the probability. Finally, we can get the efficiency of scheme P1$^\prime$ as follows:
\begin{align}\label{eq40}
Y_{P1'}=&Y_i+\frac{1}{2}(Y_{eoe/oeo}P'_{i(eoe/oeo)}\nonumber\\&+Y_{oee/eoo}P'_{i(oee/eoo)}+Y_{eeo/ooe}P'_{i(eeo/ooe)}).
\end{align}
The average fidelity of scheme P1$^\prime$ can be expressed as
\begin{align}\label{eq41}
F_{P1'}&\!=\!F_i\frac{Y_i}{Y_{P1'}}+F''_{0(eoe/oeo)}\!\frac{\frac{1}{2}Y_{eoe/oeo}\!P'_{i(eoe/oeo)}}{Y_{P1'}}
\!+\!F''_{0(oee/eoo)}\nonumber\\
&\times\!\frac{\frac{1}{2}Y_{oee/eoo}\!P'_{i(oee/eoo)}}{Y_{P1'}}\!+\!F''_{0(eeo/ooe)}\!\frac{\frac{1}{2}Y_{eeo/ooe}\!P'_{i(eeo/ooe)}}{Y_{P1'}}.
\end{align}

Comparing Eq. \eqref{eq40} with Eq. \eqref{eq29}-\eqref{eq32}, it is obvious that the efficiency $Y_{P1'}$ of scheme P1$^\prime$ is not as high as that $Y_{P1}$ of scheme $P1$, namely, $Y_{P1}\!>\! Y_{P1'}$, because we didn't include the cross combinations with residual entanglement of the second purification into the calculation. Thus, 
the parties may obtain higher fidelity in some cases at the cost of a little lower efficiency. 
For instance, when the initial fidelities of the symmetric three-photon ensembles are taken as ${^1\!}F_0\!=\!0.8$ and ${^2\!}F_0\!=\!0.6$, one can find that scheme P1$^\prime$ succeeds while scheme P1 fails, and $F'''_0\!=\!0.866\!>\! F^{t}_0\!=\!0.795$, $F_{P1'}\!=\!0.947\!>\! F_{P1}\!=\!0.898$, and $Y_{P1'}\!=\!0.510$, $Y_{P1}\!=\!0.753$. We also give the image of average fidelity $F_{P1'}$ of scheme P1$^\prime$ in the case of symmetric initial ensembles, as shown in Fig. \ref{pic12}.

\nocite{*}

\begin{thebibliography}{}
\bibitem{qt}C. H. Bennett, G. Brassard, C. Crepeau, R. Jozsa, A. Peres, and W. K. Wootters, Teleporting an unknown quantum state via dual classical and Einstein-Podolsky-Rosen channels, Phys. Rev. Lett. 70, 1895 (1993).
\bibitem{qkd1}A. K. Ekert, Quantum cryptography based on Bell’s theorem, Phys. Rev. Lett. 67, 661 (1991).
\bibitem{qkd2}C. H. Bennett, G. Brassard, and N. D. Mermin, Quantum cryptography without Bell’s theorem,  Phys. Rev. Lett. 68, 557 (1992).

\bibitem{QSDC1}G. L. Long and X. S. Liu, Theoretically efficient high-capacity
quantum-key-distribution scheme, Phys. Rev. A 65, 032302 (2002).
\bibitem{QSDC2}F. G. Deng, G. L. Long, and X. S. Liu, Two-step quantum direct communication protocol using the Einstein-Podolsky-Rosen pair block, Phys. Rev. A 68, 042317 (2003).
\bibitem{QSDC3}W. Zhang, D. S. Ding, Y. B. Sheng, L. Zhou, B. S. Shi, and G. C. Guo, Quantum secure direct communication with quantum
memory, Phys. Rev. Lett. 118, 220501 (2017).
\bibitem{QSDC4}Z. T. Qi, Y. H. Li, Y. W. Huang, J. Feng, Y. L. Zheng, and
X. F. Chen, A 15-user quantum secure direct communication
network, Light Sci. Appl. 10, 183 (2021).
\bibitem{QSDC5}Y. B. Sheng, L. Zhou, and G. L. Long, One-step quantum secure
direct communication, Sci. Bull. 67, 367 (2022).
\bibitem{QSDC6} L. Zhou and Y. B. Sheng, One-step device-independent quantum secure direct communication, Sci. China Phys. Mech.
Astron. 65, 250311 (2022).
\bibitem{DSQML}Y. B. Sheng and L. Zhou, Distributed secure quantum machine
learning, Sci. Bull. 62, 1025 (2017).
\bibitem{repeater1}H. J. Briegel, W. Dür, J. I. Cirac, and P. Zoller, Quantum repeaters: the role of imperfect local operations in quantum communication, Phys. Rev. Lett. 81, 5932 (1998).

\bibitem{qss1}M.Hillery, V. Bu\v{z}ek, and A. Berthiaume, Quantum secret sharing, Phys. Rev. A 59, 1829 (1999).
\bibitem{qss2}A. Karlsson, M. Koashi, and N. Imoto, Quantum entanglement for secret sharing and secret splitting, Phys. Rev. A 59, 162 (1999).
\bibitem{qss3}L. Xiao, G. L. Long, F. G. Deng, and J. W. Pan, Efficient multiparty quantum-secret-sharing schemes, Phys. Rev. A 69, 052307 (2004).
\bibitem{qsts1}R. Cleve, D. Gottesman, and H. K. Lo, How to share a quantum secret, Phys. Rev. Lett. 83, 648 (1999).
\bibitem{qsts2}A. M. Lance, T. Symul, W. P. Bowen, B. C. Sanders, and P. K. Lam, Tripartite quantum state sharing, Phys. Rev. Lett. 92, 177903 (2004).
\bibitem{qsts3}F. G. Deng, X. H. Li, C. Y. Li, P. Zhou, and H. Y. Zhou, Multiparty quantum-state sharing of an arbitrary two-particle state with Einstein-Podolsky-Rosen pairs, Phys. Rev. A 72, 044301 (2005).

\bibitem{quantumcomputation}M. A. Nielsen and I. L. Chuang, Quantum computation and quantum information (Cambridge University Press, Cambridge, 2000).

\bibitem{repp1}C. H. Bennett, G. Brassard, S. Popescu, B. Schumacher, J. A. Smolin, and W. K. Wootters, Purification of noisy entanglement and faithful teleportation via noisy channels, Phys. Rev. Lett. 76, 722 (1996).
\bibitem{diffepp1}D. Deutsch, A. Ekert, R. Jozsa, C. Macchiavello, S. Popescu, and A. Sanpera, Quantum privacy amplification and the security of quantum cryptography over noisy channels, Phys. Rev. Lett. 77, 2818 (1996).
\bibitem{earlymepp1}M. Murao, M. B. Plenio, S. Popescu, V. Vedral, and P. L. Knight, Multiparticle entanglement purification protocols, Phys. Rev. A 57, R4075 (1998).
\bibitem{repp2}J. W. Pan, C. Simon, and A. Zellinger, Entanglement purification for quantum communication, Nature (London) 410, 1067 (2001).




\bibitem{repp3}C. Simon and J. W. Pan, Polarization entanglement purification using spatial entanglement, Phys. Rev. Lett. 89, 257901 (2002).
\bibitem{diffepp2}W. Dür and H. J. Briegel, Multiparticle entanglement purification for graph states, Phys. Rev. Lett. 91, 107903 (2003).
\bibitem{rmepp5}Y. W. Cheong, S.W. Lee, J. Lee, and H. W. Lee, Entanglement purification for high-dimensional multipartite systems, Phys. Rev. A 76, 042314 (2007).
\bibitem{repp4}Y. B. Sheng, F. G. Deng, and H. Y. Zhou, Efficient polarization-entanglement purification based on parametric down-conversion sources with cross-Kerr nonlinearity, Phys. Rev. A 77, 042308 (2008).

\bibitem{earlymepp2}Y. B. Sheng, F. G. Deng, and H. Y. Zhou, Multipartite entanglement purification with quantum
nondemolition detectors, Eur. Phys. J. D 55, 235 (2009).
\bibitem{depp1}Y. B. Sheng and F. G. Deng, Deterministic entanglement purification and complete nonlocal Bell-state analysis with hyperentanglement, Phys. Rev. A 81, 032307 (2010).
\bibitem{depp2}Y. B. Sheng and F. G. Deng, One-step deterministic polarization-entanglement purification using spatial entanglement, Phys. Rev. A 82, 044305 (2010).
\bibitem{depp3}X. H. Li, Deterministic polarization-entanglement purification using spatial entanglement, Phys. Rev. A 82, 044304 (2010).
\bibitem{earlymepp3}Y. B. Sheng, F. G. Deng, and G. L. Long, Multipartite electronic entanglement purification with charge detection, Phys. Lett. A 375, 396 (2011).
\bibitem{rmepp4}M. Huber and M. Plesch, Purification of genuine multipartite entanglement, Phys. Rev. A 83, 062321 (2011).


\bibitem{dmepp1}F. G. Deng, One-step error correction for multipartite polarization entanglement, Phys. Rev. A 83, 062316 (2011).
\bibitem{36-pra-deng-mepp}F. G. Deng, Efficient multipartite entanglement purification with the entanglement link from a subspace, Phys. Rev. A 84, 052312 (2011).
\bibitem{dmepp2}Y. B. Sheng, G. L. Long, and F. G. Deng, One-step deterministic multipartite entanglement purification with linear optics, Phys. Lett. A 376, 314 (2012).
\bibitem{mbepp1}M. Zwerger, W. Dür, and H. J. Briegel, Measurement-based quantum repeaters, Phys. Rev. A 85, 062326 (2012).
\bibitem{mbepp2}M. Zwerger, H. J. Briegel, and W. Dür, Universal and optimal error thresholds for measurement-based entanglement purification, Phys. Rev. Lett. 110, 260503 (2013).

\bibitem{hybrid}Y. B. Sheng, L. Zhou, G. L. Long, Hybrid entanglement purification for quantum repeaters, Phys. Rev. A 88, 022302 (2013).
\bibitem{mbepp3}M. Zwerger, H. J. Briegel, and W. Dür, Robustness of hashing protocols for entanglement purification, Phys. Rev. A 90, 012314 (2014).

\bibitem{optimization-epp2}S. Krastanov, V. V. Albert, and L. Jiang, Optimized entanglement purification, Quantum 3, 123123 (2019).

\bibitem{optimization-epp1}F. Rozpeędek, T. Schiet, D. Elkouss, A. C. Doherty, and S. Wehner, Optimizing practical entanglement distillation, Phys. Rev. A 97, 062333 (2018).

\bibitem{13-oe-diffepp}L. Zhou, W. Zhong, Y. B. Sheng, Purification of the residual entanglement, Optics Express. 28, 383499 (2020).

\bibitem{dur}F. Riera-Sàbat, P. Sekatski, A. Pirker, and W. Dür, Entanglement-assisted entanglement purification, Phys. Rev. Lett. 127, 040502 (2021)

\bibitem{mbmepp1}P. S. Yan, L. Zhou, W. Zhong, and Y. B. Sheng, Measurement-based logical qubit entanglement purification, Phys. Rev. A 105, 062418 (2022).
\bibitem{mbmepp2}P. S. Yan, L. Zhou, W. Zhong, and Y. B. Sheng, Feasible
measurement-based entanglement purification in linear optics, Opt. Express 29, 9363 (2021).
\bibitem{mbmepp3}P. S. Yan, L. Zhou, W. Zhong, and Y. B. Sheng, Measurement-based entanglement purification for entangled coherent states, Front. Phys. 17, 21501 (2022).

\bibitem{experiment1}J. W. Pan, S. Gasparoni, R. Ursin, G. Weihs, and A. Zeilinger,
Experimental entanglement purification of arbitrary unknown
states, Nature (London) 423, 417 (2003).

\bibitem{experiment2}X. M. Hu, C. X. Huang, Y. B. Sheng, L. Zhou, B. H. Liu, Y. Guo, C. Zhang, W. B. Xing, Y. F. Huang, C. F.
Li, and G. C. Guo, Long-distance entanglement purification for quantum communication, Phys. Rev. Lett. 126, 010503
(2021).
\bibitem{experiment3}S. Ecker, P. Sohr, L. Bulla, M. Huber, M. Bohmann, and R.
Ursin, Experimental single-copy entanglement distillation,
Phys. Rev. Lett. 127, 040506 (2021).

\bibitem{experiment4}C. X. Huang, X. M. Hu, B. H. Liu, L. Zhou, Y. B. Sheng,
C. F. Li, and G. C. Guo, Experimental one-step deterministic
polarization entanglement purification, Sci. Bull. 67, 593
(2022).
\bibitem{experiment5}S. Ecker, P. Sohr, L. Bulla, R. Ursin, and M. Bohmann, Remotely establishing polarization entanglement over noisy
polarization channels, Phys. Rev. Applied 17, 034009
(2022).
\bibitem{experiment6}R. Reichle, D. Leibfried, E. Knill, J. Britton, R. B. Blakestad, J. D.
Jost, C. Langer, R. Ozeri, S. Seidelin, and D. J. Wineland, Experimental purification of two-atom entanglement, Nature
443, 838 (2006).
\bibitem{experiment7}N. Kalb, A. A. Reiserer, P. C. Humphreys, J. J. W. Bakermans, S.
J. Kamerling, N. H. Nickerson, S. C. Benjamin, D. J. Twitchen, M. Markham, and R. Hanson, Entanglement distillation between solid-state quantum network nodes, Science 356, 928 (2017),

\bibitem{experiment8}H. X. Yan, Y. P. Zhong, H. S. Chang, A. Bienfait, M. H. Chou,
C. R. Conner, E. Dumur, J. Grebel, R. G. Povey, and A. N.Cleland, Entanglement purification and protection in a superconducting
quantum network, Phys. Rev. Lett. 128, 080504
(2022).
\bibitem{review}P. S. Yan, L. Zhou, Y. B. Sheng, and W. Zhong, Advances in quantum entanglement purification, Sci. China Phys. Mech. Astron. 66, 250301 (2023).

\bibitem{quantum-repeaters-with-multiplexed-memory1}O. A. Collins, S. D. Jenkins, A. Kuzmich, and T. A. B. Kennedy, Multiplexed memory-insensitive quantum repeaters, Phys. Rev. Lett. 98, 060502 (2007).
\bibitem{quantum-repeaters-with-multiplexed-memory2}M. Zwerger, A. Pirker, V. Dunjko, H. J. Briegel, and W. Dür, Long-range big quantum-data transmission, Phys. Rev. Lett. 120, 030503 (2018).

\bibitem{qnd-kerr}K. Nemoto and W. J. Munro, Nearly deterministic linear optical Controlled-NOT gate, Phys. Rev. Lett. 93,250502 (2004).

\bibitem{PhysRevA.68.064303} E. Knill, Bounds on the probability of success of postselected nonlinear sign shifts implemented with linear optics, Phys. Rev. A 68, 064303 (2003).




\end{thebibliography}

\end{document}